\documentclass[final,1p,times,preprint]{elsarticle}

\usepackage{amssymb}
\usepackage{amsmath}
\usepackage{color}
\usepackage{url}
\usepackage{booktabs}
\usepackage{multirow}
\usepackage{bbm}
\usepackage{multicol}
\usepackage{comment}
\usepackage{placeins}
\usepackage{xparse} 
\usepackage{hyperref}

\newcommand{\bm}[1]{\boldsymbol{#1}}

\newcommand{\bu}{\bm{u}}

\newcommand{\bF}{\bm{F}}

\newcommand{\bS}{\bm{S}}
\newcommand{\bI}{\bm{I}}

\newcommand{\bq}{\bm{q}}
\newcommand{\bQ}{\bm{Q}}

\newcommand{\btheta}{\bm{\theta}}

\newcommand{\R}{\mathbb{R}}

\newcommand{\pp}[2]{\frac{\partial #1}{\partial #2}}
\newcommand{\dd}[2]{\frac{d #1}{d #2}}
\newcommand{\ol}[1]{\overline{#1}}

\journal{arXiv}

\begin{document}


\title{OGF: An Online Gradient Flow Method for Optimizing the Statistical Steady-State Time Averages of Unsteady Turbulent Flows} 
\date{}

\author[1]{Tom Hickling\corref{aaa}}
\author[2]{Jonathan F. MacArt}
\author[1]{Justin Sirignano}
\author[1]{Den Waidmann}
\address[1]{Mathematical Institute, University of Oxford, UK.}
\address[2]{Department of Aerospace and Mechanical Engineering, University of Notre Dame, USA.\vspace{-22pt}}
\fntext[fn1]{Email: \texttt{tom.hickling@maths.ox.ac.uk}. Corresponding Author. (Author order is alphabetical.)}
\fntext[fn2]{Email: \texttt{jmacart@nd.edu}}
\fntext[fn3]{Email: \texttt{justin.sirignano@maths.ox.ac.uk}}
\fntext[fn4]{Email: \texttt{den.waidmann@maths.ox.ac.uk}}

\begin{abstract}
Turbulent flows are chaotic and unsteady, but their statistical distribution converges to a statistical steady state. Engineering quantities of interest typically take the form of time-average statistics such as $ \frac{1}{t} \int_0^t f ( u(x,\tau; \theta) ) d\tau \overset{t \rightarrow \infty}{\rightarrow} F(x; \theta)$, where $u(x,t; \theta)$ are solutions of the Navier--Stokes equations with parameters $\theta$. Optimizing over the time-averaged statistic $F(x; \theta)$ has many engineering applications including geometric optimization, flow control, and closure modeling. However, optimizing $F(x; \theta)$ is non-trivial and currently remains an open challenge, as existing computational approaches are incapable of scaling to physically representative mesh resolutions, which can require more than $\mathcal{O}(10^7)$ degrees of freedom (number of PDE variables $\times$ mesh points). The fundamental obstacle is the chaoticity of turbulent flows: gradients calculated with the adjoint method diverge exponentially as $t \rightarrow \infty$.

We develop a new online gradient-flow (OGF) method that is scalable to large degree-of-freedom systems and enables optimizing for the steady-state statistics of chaotic, unsteady, turbulence-resolving simulations. The method forward-propagates an \emph{online} estimate for the gradient of $F(x; \theta)$ while simultaneously performing online updates of the parameters $\theta$. A key feature is the fully online nature of the algorithm to facilitate faster optimization progress and its combination with an online finite-difference estimate to avoid the divergence of gradients due to chaoticity. Unlike standard finite-difference estimators, the online estimator requires a careful decomposition of the objective function gradient into the product of an error term and an instantaneous gradient term estimated with independent realizations of the chaotic dynamics. The online gradient flow can be viewed as a form of stochastic gradient descent for optimizing chaotic dynamics, where a noisy online estimate is calculated for the direction of steepest descent and the parameter is continuously updated in the direction of this noisy estimate. 

The convergence of the method can be accelerated using large minibatches of independent, parallel simulations. We demonstrate the proposed OGF method for optimizations over three chaotic ordinary and partial differential equations: the Lorenz-63 equation, the Kuramoto--Sivashinsky equation, and Navier--Stokes solutions of compressible, forced, homogeneous isotropic turbulence. In each case, the OGF method successfully reduces the loss based on $F(x; \theta)$ by several orders of magnitude and accurately recovers the optimal parameters.  
\end{abstract}
 
\maketitle

\section{Introduction}
\subsection{Background}
Some examples of turbulence-resolving simulation methods in industrially relevant scenarios include turbulence-resolving simulations of automotive~\cite{Ashton2024} and aerospace~\cite{Goc2021} external aerodynamics and combustion~\cite{Benard2019}.
Engineering quantities of interest from these simulations typically take the form of time-averaged statistics calculated over a statistical steady state of the turbulent flow field.
Optimizing over these steady-state statistics is highly desirable and could lead to improved geometries, better flow controllers, and more accurate models for unresolved physics. However, to the authors' knowledge, there are currently no existing approaches that are capable of accurately optimizing over a large time horizon while still being able to scale to the large numbers of degrees of freedom (at least $\mathcal{O}(10^7)$, given by the number of PDE variables $\times$ grid points) that are characteristic of engineering simulations.

Adjoint methods are typically the tool of choice for equation-constrained optimization. However, for turbulent flows, adjoint methods are only capable of optimizing over relatively short time intervals and diverge exponentially as $t\rightarrow\infty$ due to the chaotic nature of the flow (we demonstrate this in Section~\ref{sec:chaos_challenges}, see also~\citet{LEA2000}).
In addition, they require storing the entire solution history for use in the backward pass, which can result in excessive memory/storage requirements---potentially in the range of terabytes to petabytes depending on the time horizon.
In the remainder of this introduction, we explain the challenges of optimizing over the statistical steady state of chaotic systems, discuss existing techniques to address these challenges, and outline the present contributions.

\subsection{The challenge of optimization in chaotic systems}\label{sec:chaos_challenges}
Consider the $\ell_2$ loss function
\begin{equation}\label{eq:loss}
    J(\theta) = \left(\lim_{t\rightarrow\infty}\left[\frac{1}{t}\int_0^t f(u(\tau; \theta))\ d\tau\right] - F^\ast\right)^2,
\end{equation}
where $f(u)$ is a function of the ODE/PDE solution $u$, $F^*$ is the target value of $f$, and $\theta$ is a multi-dimensional parameter governing the dynamical system ${du}/{dt} = R(u; \theta)$. We wish to select the $N_p$-dimensional parameter $\theta \in \mathbb{R}^{N_p}$ to minimize the distance between the time average of $f(u(\tau; \theta))$ and the target $F^*$. This requires optimizing over the dynamics of $u$, which are assumed to be chaotic and unsteady but \emph{ergodic}. Ergodicity means that the statistical distribution of $u(t; \theta)$---in particular time-averages such as $\frac{1}{t}\int_0^t f( u(\tau; \theta)) \ d\tau $ ---converges in the limit $t \rightarrow \infty$. 
Ergodicity is a basic assumption in the theory of turbulence~\cite{E2001} for which there is much practical evidence (some explicit tests of the assumption are detailed in~\citet{Galanti2004} and~\citet{Djenidi2013}). In simpler chaotic systems, ergodicity has been proven as log as the initial conditions are sufficiently close to the attractor---see~\citet{Ghys2013} for a discussion of the Lorenz-63 system~\cite{Lorenz1963}.

A natural question is whether gradient descent can be used to minimize $J(\theta)$. Using the chain rule, the derivative of $J$ with respect to the parameter $\theta$ can be written as
\begin{equation} \label{dJdtheta0}
    \pp{J}{\theta} = \underbrace{2 \left( 
    \lim_{t\rightarrow\infty}\left[\frac{1}{t}\int_0^t f(u(\tau; \theta))\ d\tau\right] 
    - F^* \right)}_{\dfrac{\partial J}{\partial\langle f(u(\cdot;\ \theta) \rangle}}
    \underbrace{\pp{}{\theta} \left( \lim_{t\rightarrow\infty}\left[\frac{1}{t}\int_0^t f(u(\tau; \theta))\,d\tau\right] \right)}_{\displaystyle\dfrac{\partial}{\partial\theta}\langle f(u(\cdot;\ \theta)) \rangle},
\end{equation}
where $\langle f(u(\cdot\ ;\theta) \rangle = \lim_{t\rightarrow\infty}\left[\frac{1}{t}\int_0^t f(u(\tau; \theta))\ d\tau\right]$. 

At this point, from a practical perspective, one might wish to interchange the outer derivative and inner derivative in the second term in (\ref{dJdtheta0}), thereby obtaining $\langle \pp{}{\theta} f(u(\cdot\ ; \theta))\rangle$. $\pp{u}{\theta}$ can then be evaluated using the adjoint method or a forward sensitivity method. Unfortunately, this requires the assumption that the derivative $\pp{}{\theta}$ and the limit $\lim_{t \rightarrow \infty}$ commute, which is not true in a chaotic system~\cite{Blonigan2018}. This fundamentally leads to extreme difficulty in optimizing over the steady-state statistics of a chaotic system.
In fact, if one attempted to numerically evaluate $\langle \pp{}{\theta} f(u(\cdot\ ; \theta))\rangle$, it would diverge (growing to $\pm \infty$) instead of converging to $\pp{}{\theta} \langle f(u(\cdot\ ; \theta))\rangle$. Specifically, calculating $\frac{\partial}{\partial \theta}f(u(\tau; \theta))$ will be numerically unstable (i.e., diverging as $t \rightarrow \infty$) and subsequently time-averaging over $[0,t]$ will not converge to $\frac{\partial}{\partial \theta} \langle f(u(\cdot\ ; \theta))\rangle$. Conventional unsteady adjoint and forward sensitivity equations for time-averaged quantities are therefore numerically unstable for large times~\cite{Wang2014, Chung2022}.

\begin{figure}[t!]
    \centering
    \includegraphics[width=0.8\linewidth]{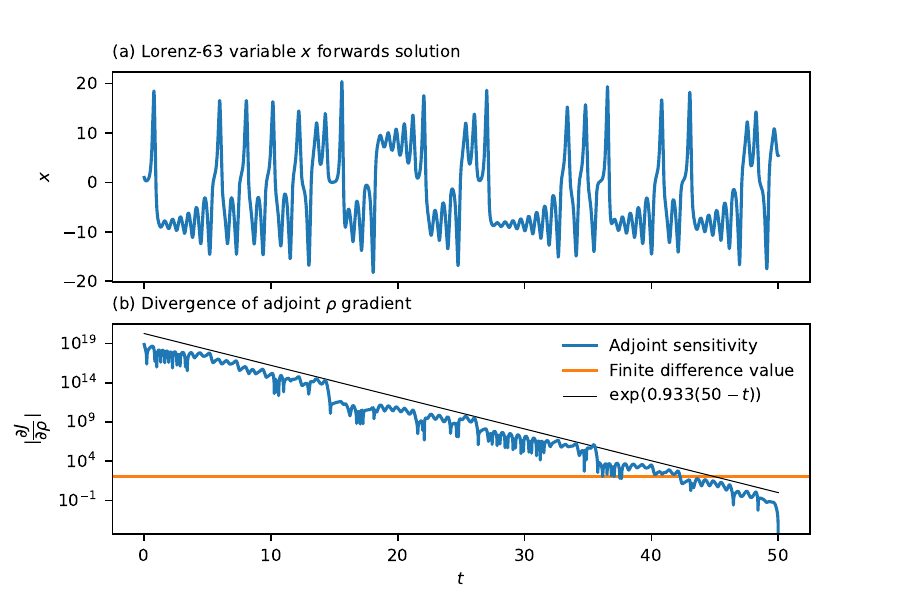}
    \caption{Adjoint calculation of $\pp{J_{\tau_{opt}}}{\rho}$ for the Lorenz-63 system with the loss function \eqref{eq:toy_loss_function} and $\tau_{opt} = 50$. The value of 0.933 for the Lyapunov exponent is from \citet{LEA2000}.}
    \label{fig:adj_blowup}
\end{figure}

For the practical consequences of this, consider the Lorenz-63 ordinary differential equation (ODE) system~\cite{Lorenz1963}, a reduced-order model for natural convection given by
\begin{equation}\label{eq:lorenz63}
\dd{}{t}\begin{Bmatrix}
    x\\y\\z
\end{Bmatrix} = 
\begin{Bmatrix}
    \sigma (y - x)\\ x(\rho - z) - y\\ xy - \beta z
\end{Bmatrix}.
\end{equation}
This system is well-known to exhibit chaotic behavior for the classical parameters 
$\theta = \lbrace \rho,\ \sigma,\ \beta \rbrace^\top = \lbrace 28,\ 10,\ 8/3\rbrace^\top$. Denoting the forwards-in-time (hereafter referred to as simply ``forwards'') ODE solution as $u = \lbrace x,\ y,\ z\rbrace^\top$ and its backwards-in-time adjoint as 
\begin{equation}
\hat{u} = \pp{J_{\tau_{opt}}}{u},
\end{equation}
the adjoint describes the gradient of the finite-time average loss function $J_{\tau_{opt}}$ over the time interval $[0, \tau_{opt}]$ with respect to the solution at the current time.
The adjoint equation for the Lorenz-63 system, with an explicit Euler discretization, is given by
\begin{equation}\label{eq:lorenz_adjoint}
\hat{u}_{n-1} =
\hat{u}_{n}+
\Delta t
\begin{bmatrix}
        -\sigma & \sigma & 0\\
    (\rho - z_{n-1}) & -1 & -x_{n-1}\\
    y_{n-1} & x_{n-1}& -\beta
\end{bmatrix}
\hat{u}_{n}
+ 
\pp{J_{\tau_{opt}}}{u_{n-1}}.
\end{equation}
The gradient of $J_{\tau_{opt}}$ with respect to the parameters $\theta$ can be calculated using the adjoint solution as
\begin{equation}\label{eq:lorenz_sensitivity}
    \pp{J_{\tau_{opt}}}{\theta} = \sum_n \hat{u}_n^{\top} \left.\pp{u_{n}}{\theta}\right|_{n-1}.
\end{equation}
A gradient calculation using the adjoint method \eqref{eq:lorenz_sensitivity} is shown in Fig.~\ref{fig:adj_blowup}.
This calculation uses a loss function
\begin{equation}\label{eq:toy_loss_function}
    J_{\tau_{opt}}(\theta) = 
    \left(\frac{\langle x^2\rangle_{\tau_{opt}}}{\langle x^2 \rangle^\ast}\right)^2 + 
    \left(\frac{\langle y^2\rangle_{\tau_{opt}}}{\langle y^2 \rangle^\ast}\right)^2 + 
    \left(\frac{\langle z^2\rangle_{\tau_{opt}}}{\langle z^2 \rangle^\ast}\right)^2,
\end{equation}
where $\langle x^2 \rangle_{\tau_{opt}} = \frac{1}{\tau_{opt}}\int_{0}^{\tau_{opt}} x(t)^2 dt$, $\langle x^2 \rangle^\ast = 67.84$, $\langle y^2 \rangle^\ast = 84.02$, and $\langle z^2 \rangle^\ast = 689.0$ are the \emph{a priori} known values for the statistics with the classical Lorenz parameters.

While the forward solution (Fig.~\ref{fig:adj_blowup}a) is bounded, the backward-in-time adjoint solution (Fig.~\ref{fig:adj_blowup}b) is not---the gradient diverges exponentially at a rate that agrees with the system's largest Lyapunov exponent~\cite{Wang2014}.
Also shown in Fig.~\ref{fig:adj_blowup}b is a standard finite difference estimate for the true gradient $\pp{J}{\rho}$, computed as
\begin{equation}\label{eq:simple_FD}
    \pp{J}{\rho} \approx \frac{J(\theta_{\rho+\varepsilon}) - J(\theta_{\rho-\varepsilon})}{2\varepsilon},
\end{equation}
with perturbed parameter vector $\theta_{\rho\pm\varepsilon} = \lbrace \rho \pm \varepsilon,\ \sigma,\ \beta\rbrace$ and $\varepsilon=1.0$.
For this example, the perturbed loss estimates are calculated over a 1,000-time-unit horizon (randomly initialized from the ergodic state)  with a 100,000-trajectory minibatch. 
The adjoint gradient does not stay in the neighborhood of the true gradient for any reasonable amount of time and so would be unusable for optimization.

\subsection{Existing approaches for the optimization of chaotic systems}\label{sec:existing_chaos_optimizers}
Many approaches have been proposed to enable gradient calculations and optimization of time-averaged statistics of chaotic systems.
We categorize existing methods as follows: short time horizon adjoints \cite{LEA2000}, Ruelle's linear response formula \cite{Ruelle-1997,Ruelle-2003}, artificial dissipation added to the backward adjoint solution~\cite{Blonigan-2012}, reformulating the forward problem so that it is more amenable to adjoint analysis~\cite{Lasagna2018,Burton2024,Blonigan2014a}, shadowing trajectory derived approaches~\cite{Blonigan2017}, reinforcement learning~\cite{Bae-2022,Vinuesa-2022}, and finite-difference based methods~\cite{LEA2000}.

Ensemble (adjoint) methods, first proposed by \citet{LEA2000}, simulate a set of sample trajectories started from different initial conditions over a small time horizon, which limits the exponential growth of unstable modes in the adjoint.
However, due to the small time horizon, these approaches are by definition unable to optimize over time-averaged statistics---the gradient cannot include information about time scales that are longer than the time horizon for which it was calculated.
\citet{Liu2023} and \citet{Liu-2025} applied ensemble adjoint methods to control 2D laminar and 3D turbulent flows around cylinders and airfoils. In \cite{Sirignano2020,MacArt2021,Sirignano2022,Hickling2024},
adjoint methods have been applied to optimize subgrid-scale (SGS) closure models for LES. 
Other techniques leveraging shorter time horizons also exist, for example the penalty-based method of \citet{Chung2022}. While it is effective in reducing the loss for a 3D Kolmogorov flow with forcing, the method does not guarantee that the computed time evolution of the solution fulfills the Navier--Stokes equations.

Ruelle's linear response formula \cite{Ruelle-1997, Ruelle-2003} offers another way of computing gradients. \citet{Eyink-2004} show that a reformulation of Ruelle's formula is equivalent to the ensemble method of \citet{LEA2000}. \citet{Eyink-2004} apply the formula to the Lorenz system and numerically establish its convergence rate in the number of samples.
\citet{Chandramoorthy2019} found that, even under optimistic assumptions of exponential decay of the bias in the gradient estimate, the computational cost of these methods made them infeasible for practical applications.

Including artificial dissipation in the backwards adjoint system can be used to prevent exponentially growing gradients and to maintain a stable energy budget.
Various techniques for doing this have been explored \cite{Blonigan-2012,Talnikar2016,Ashley-2019,Garai2021,Bhatia2019}.
The methods are effective at stabilizing the adjoint calculation; however, adding artificial dissipation to the adjoint equation introduces a discrepancy with the true dynamics of the system (e.g., the Navier--Stokes equations); therefore, the resulting adjoint does not calculate the correct gradient for the objective function. Thus, while adding artificial dissipation can prevent numerical instability, it leads to an inconsistency between the adjoint equation and the forward dynamics which one is interested in optimizing over \cite{Garai2021}.

It is also possible to modify the governing equations so that an equivalent ``steady" problem is solved without a time domain, which enables the use of standard adjoint methods. Methods for doing this include finding unstable periodic orbits~\cite{Lasagna2018}, transforming the time domain into the frequency domain \cite{Burton2024}, or solving the Fokker--Planck equation for the probability density function of the solution~\cite{Blonigan2014a,Yang2023}.
Although these approaches effectively bypass the issue of a diverging unsteady adjoint, their \emph{forwards} solution is often computationally challenging, so they have not (to our knowledge) been applied to high degree-of-freedom numerical systems that are routinely solved by standard finite-difference numerical methods.

\citet{Wang2013} proposed forward and adjoint sensitivity methods based on the inversion of the so-called shadow operator and applied it to the Lorenz-63 system. 
Inversion of the shadowing operator is expensive, which has led to the development of the least squares shadowing method (LSS)~\cite{Wang2014,Wang2014a}, which obtains the gradient of the loss function by computing a non-diverging shadow trajectory as the solution of a Karush--Kuhn--Tucker (KKT) system over the full forward time horizon.
The method has been analyzed for homogeneous isotropic turbulence  \cite{Blonigan-2014}, a 2D airfoil  \cite{Blonigan2018} and the Kuramoto--Sivashinsky equation (KSE)  \cite{Blonigan2014}. Due to the associated cost of solving the KKT system, multiple variants have been suggested, such as a multigrid method \cite{Blonigan-Wang-2014}, simplified LSS \cite{Chater-2017}, multiple-shooting shadowing \cite{BLONIGAN2018447}, and non-intrusive-LSS (NILSS) \cite{Ni2017,Blonigan2017}.
The cost of NILSS (in the adjoint formulation) is independent of the parameter dimension but scales with the number of positive Lyapunov exponents. Hence, turbulent flows with high Reynolds numbers are challenging to optimize over using these methods. Other applications include a minimal flow unit~\cite{Blonigan2017} and weakly turbulent flow over a cylinder~\cite{Ni_2019_turb}.

Reinforcement learning (RL) techniques have found use in turbulent flows for training deep-learning closure models or flow controllers. 
RL methods are applicable to a broad category of problems, as they do not require knowledge of the underlying system and in practice can be used on top of existing code bases. The effectiveness of RL has been demonstrated for LES wall-modeling LES \cite{Bae-2022,Zhou2022,Vadrot-2023}, closure modeling~\cite{Novati2021}, and control  \cite{Gazzola2014,Verma2018,Bucci2019,Vinuesa-2022,Liu2023,Chatzimanolakis2024}. 
The generality and flexibility of RL may come with a higher computational cost: for example, \citet{Liu2023} found RL to be more computationally costly than gradient-based optimization for control of unsteady flow around a cylinder. Our proposed method directly estimates the gradient of the objective function, while RL typically simultaneously estimates a ``critic" for the value function and an ``actor" for the control, where the actor is trained using the critic. Typical RL methods will optimize over the discounted future error (e.g., an objective function $J(\theta) = \int_0^{\infty} \mathbf{e^{-rt}} ( f(u_t; \theta) - F )^2 dt $) with a discount factor $e^{-rt}$ (where the constant $r > 0$) while our method can directly optimize over the time average of the solution (i.e., the moments of the statistical steady-state distribution of chaotic dynamics) as in the objective function (\ref{eq:loss})). Therefore, RL and our proposed method are complementary in the sense that they can optimize different classes of objective functions. 

Finally, finite-difference methods can provide accurate gradients to optimize over time-aver\-aged quantities in chaotic systems. A discussion of their application to the Lorenz system is found in \citet{LEA2000}, who found that  finite-difference gradients provide adequate estimates of the system's response to a change in parameters, provided that the perturbed and unperturbed quantities have been simulated long enough. In addition, one needs to simulate an extra trajectory for each parameter to be optimized over. Modern parallel machines are able to compute those for a small set of parameters, but many optimization iterations are often required for the gradient-descent algorithm to converge, which requires re-simulating the system many times to calculate the finite-difference gradients. This process can easily be computationally prohibitive for systems of engineering interest.

\subsection{Contributions of this paper}
We develop a new online gradient-flow (OGF) method to optimize over the statistical steady-state of chaotic systems. The method is able to scale to high degree-of-freedom (i.e. large numbers of mesh points) simulations of PDEs.  

The method forward-propagates an \emph{online}, finite-difference estimate for the gradient  $\nabla_{\theta} F(x; \theta)$, where
\begin{equation}
F(x; \theta) = \lim_{\tau \rightarrow \infty} \left[\frac{1}{\tau} \int_0^\tau f ( u(x, \tau; \theta) ) dt\right],
\end{equation}
while simultaneously performing online updates of the parameter $\theta$. Key features include the fully online nature of the algorithm, which facilitates faster optimization progress, and its combination with a finite-difference estimate to avoid the divergence of gradients due to chaoticity. 
Unlike in standard finite-difference estimators, the online estimator requires a careful decomposition of the objective function gradient into the product of an error term and an instantaneous gradient term estimated with independent realizations of the chaotic dynamics. The convergence of the method can be accelerated using large minibatches of independent parallel simulations. 

We evaluate the OGF method for several chaotic ODE and PDE systems. These include multi-parameter optimizations over the Lorenz-63 equation and the Kuramoto--Sivashinsky equation as well as optimization of a single parameter in a forced homogeneous isotropic turbulence  (HIT) simulation of the compressible Navier--Stokes equations. The online gradient-flow method successfully reduces the loss by multiple orders of magnitude and accurately recovers the target parameters in each of these numerical examples. The method is generally robust to the choice of hyperparameters, as long as the learning rate is small enough and decays according to the usual convergence requirements of gradient descent methods.

There are several key advantages of the OGF method. First, it is able to scale to large degree-of-freedom systems; we demonstrate this with a compressible HIT example containing $8.4 \times 10^7$ degrees of freedom. Second, its convergence can be accelerated by increasing the number of independent minibatch simulations, which can be fully parallelized. Third, due to its online nature, the gradient-flow method does not require storing solution trajectories for backward propagation or target data, which is infeasible for long time horizons. Finally, a na\"ive finite-difference estimator for the gradient $\nabla_{\theta} J(\theta)$ would be an iterative method: at each iteration a long simulation would be run to estimate the gradient and then a single gradient descent step would be taken. This can be computationally expensive, since each optimization iteration requires a long simulation. In contrast, the online gradient-flow algorithm continuously updates the parameter using an online estimate of the gradient, which itself is also updated in parallel, which allows the algorithm to continually make optimization progress. 

The proposed online optimization method can be viewed as a form of stochastic gradient descent for optimizing chaotic dynamics, where a noisy online estimate is calculated for the direction of steepest descent and the parameter is simultaneously updated in the direction of this noisy estimate. 

The paper is organized as follows.
In Section~\ref{s2} we describe the online gradient flow methodology.
We apply the method to a series of increasingly more challenging (in terms of complexity, chaoticity, and computational cost) systems: the Lorenz-63 system of ODEs~\cite{Lorenz1963} in Section~\ref{sec:lorenz}, the Kuramoto--Sivashinsky PDE~\cite{Kuramoto-1976,Kuramoto1978,Sivashinsky1977} in Section~\ref{sec:kse}, and compressible forced homogeneous isotropic turbulence in simulations with up to $8.4\times10^7$ degrees of freedom in Section~\ref{s4}.
Finally, in Section~\ref{sec:hyperparameter_sensitivity} we examine the hyperparameter sensitivity of the methodology across these applications and illustrate its general robustness.

\section{An online gradient flow (OGF) method for optimizing over chaotic systems} \label{s2}
In this section we describe the online gradient flow algorithm. 
Let $\Omega$ be an open subspace of $\R^{d}$, where $d$ is the spatial dimension.
The OGF algorithm optimizes over the parameters $\theta$ of an ergodic PDE of the form
\begin{equation} \label{Eq:GovEq}
    \pp{u}{t} = R(u;\theta),
\end{equation}
where $R$ is a partial differential operator involving spatial derivatives on the space $\Omega$. We denote a solution to (\ref{Eq:GovEq}) at location $x\in\Omega$, time $t$, and with a particular set of parameters $\theta$ as $u(x, t; \theta)$ and its initial condition as $u(x, 0; \theta) = u_0(x)$.

Denoting the time average of a generic variable $\phi(x, t )$ as
\begin{equation}
    \Big\langle\phi(x, \cdot)\Big\rangle = \lim_{\tau\rightarrow\infty}\left[
        \frac{1}{\tau}\int_{0}^{ \tau} \phi(x, t)\ dt
    \right],
\end{equation} 
our objective is to minimize the time-averaged loss function $J$, given by
\begin{equation}\label{eq:loss_with_f}
    J(\theta) := \int_\Omega \left(\Big< f\big(u(x, \cdot; \theta)\big)\Big>  -  F^{*}(x) \right)^2 dx.
\end{equation}
Here, $f$ denotes an arbitrary function of the variables we seek to optimize, and $F^{*}(x)$ are the target statistics for $f(u(x, t;\theta)$). That is, we wish to select the parameters $\theta$ such that the time average $\Big< f\big(u(x, \cdot; \theta)\big)\Big>$ and the target data $F^{*}(x)$ are as close as possible in the $L^2$ norm. 

The method can be easily further generalized to a function $f(u,x)$ which also varies with the spatial coordinate $x$. For example, for optimizing over a set of sparse point values, $f$ would be a sum of Dirac delta functions centered at each target point location. 
The extension of the method to losses based on the squared error of multiple solution variables (as used in Section~\ref{sec:lorenz}) is also straightforward.
For convenience, we will often write $\langle u(x, \cdot; \theta) \rangle$ as $\langle u(\cdot)\rangle$.

\subsection{Continuous-time gradient flow with an online estimator for the gradient}
Consider the loss function \eqref{eq:loss_with_f}.
For gradient-based optimization algorithms, the key quantity one needs to compute is the gradient $\nabla_{\theta}J(\theta)$. By the chain rule, the $i^\mathrm{th}$ partial derivative is
\begin{equation}
    \pp{J}{\theta_{i}} =\int_{\Omega}  2\Big( \langle  f(u(x, \cdot\,; \theta))\rangle  -  F^{*}(x)  \Big) 
    \pp{}{\theta_{i}} \Big\langle f(u(x, \cdot\,; \theta)) \Big\rangle\ dx, 
\end{equation}
for parameters $\theta \in \mathbb{R}^{N_p}$ (i.e., we are optimizing over $N_p$ parameters). The partial derivative with respect to $\theta_{i}$ of the time-average is especially challenging to calculate due to the chaotic dynamics. As explained earlier, it is not possible to simply interchange the partial derivative and the limit $t \rightarrow \infty$ in order to directly differentiate the dynamics $\pp{u}{t} = R(u;\theta)$. 

Our proposed method updates the parameters $\theta$ \emph{online} while simultaneously simulating the solution to the PDE (\ref{Eq:GovEq}). The parameters $\theta$ are updated as a continuous gradient flow $\theta(t)$ satisfying an ODE which is coupled with the PDE dynamics:
\begin{align}
    \dd{\theta}{t} &= - \alpha(t)\ G \big{(} u_{i,j}(t), F^{*}; \theta(t) \big{)}, \label{eq:grad_flow} \\
    \pp{}{t}\Big( u_{i,j}(t)\Big) &= R \big{(} u_{i,j}(t);\theta^{({i,j})}(t) \big{)}, \label{eq:grad_flow_PDE}
\end{align}
where $\alpha(t)$ is a learning rate and $u_{i,j}(t)$ are \emph{independent} solutions to the PDE. By independent, we mean that their initial conditions are independent random variables. The number of PDEs $u_{i,j}(t)$ which will be solved is $3N_p$, with $i = 0,  1, \ldots, N_p-1$ (equal to the number of parameters) and $j = 0,1, 2$.
$\theta^{({i,j})}(t)$ is a perturbation around $\theta(t)$ defined as
\begin{align}
\theta^{({i,0})}(t) &= \theta(t) , \notag \\
\theta^{({i,1})}(t) &= \theta(t) + \varepsilon_i e_i, \notag \\
\theta^{(i,2)}(t) &= \theta(t) - \varepsilon_i e_i,
\end{align}
where $\varepsilon_i > 0$ is the perturbation for parameter $i$, and $e_{i} \in \mathbb{R}^d$ is the $i^\mathrm{th}$ standard unit vector, with its $k^\mathrm{th}$ element given by
\begin{equation}
    e_{i}[\ k\ ] = \delta_{ik},\quad\text{where}\quad\delta_{ik} = 
    \begin{cases}
        1 & \text{if}\quad i = k,\\
        0 & \text{otherwise}.
    \end{cases}
\end{equation}
Similarly, we define $u_{i,j}(t;\theta)$ as independent solutions to the original PDE (\ref{Eq:GovEq}):
\begin{equation} \label{Eq:GovEq2}
    \pp{}{t}\Big(u_{i,j}(t;\theta)\Big) = R(u_{i,j}(t;\theta);\theta).
\end{equation}

A subtle but crucial distinction should be highlighted between the \emph{online} PDE dynamics in (\ref{eq:grad_flow_PDE}) and the original PDE dynamics in (\ref{Eq:GovEq}) or (\ref{Eq:GovEq2}). The PDE solution $u_{i,j}(t; \theta)$ in (\ref{Eq:GovEq2}) is a function of a fixed parameter $\theta$. Conversely, the PDE evolution of $u_{i,j}(t)$ in \eqref{eq:grad_flow_PDE} is governed by a continuously evolving parameter $\theta(t)$, where the parameter $\theta(t)$ is trained in parallel via the online gradient flow \eqref{eq:grad_flow}. 
The evolution of $u_{i,j}(t)$ is therefore a function of the training path of $\theta(t)$ over $\tau \in [0,t]$ (and not just the PDE RHS function $R$).

We would like to construct a function $G$ which is an asymptotically unbiased estimator of $\nabla_{\theta}J$; that is,
\begin{equation}\label{eq:bias}
    \big\langle G(u_{i,j}(\cdot), F^\ast; \theta) \big\rangle = \pp{J}{\theta} = \int_\Omega 2 \Big(\langle 
        f(u(x,\cdot\,;\theta)) \rangle - F^{*}(x) \Big)  
    \pp{}{\theta} \Big\langle
    f(u(x,\cdot\,;\theta)) \Big\rangle\ dx.
\end{equation}
Consequently, \emph{on average}, the RHS of the ODE for $\theta(t)$ in (\ref{eq:grad_flow}) points in the direction of steepest descent, minimizing the objective function $J(\theta)$. (\ref{eq:grad_flow}) can be viewed as a stochastic gradient descent (SGD) method where, due to the chaotic dynamics, $G(u_{i,j}(t), F^{*}; \theta)$ fluctuates around its mean $\nabla_{\theta}J( \theta )$. Although these fluctuations are due to (deterministic) chaos, their effect from an optimization perspective is similar to the noise from randomly selecting data samples in SGD. The key difference between classical SGD and the online gradient flow equation (\ref{eq:grad_flow}) is that, in the latter, the noise is correlated in time, while in classical SGD the noise is i.i.d.\ (independent and identically distributed) due to the randomly selected data samples. 

In particular, the RHS of (\ref{eq:grad_flow}) can be decomposed into a direction of steepest descent and a fluctuation/noise term:
\begin{equation}\label{eq:grad_flowDecomposition}
    \dd{\theta}{t} = \underbrace{- \alpha(t) \nabla_{\theta} J(\theta(t))}_{\textrm{Gradient descent in direction of steepest descent } } + \underbrace{ \alpha(t)\bigg{(}   \nabla_{\theta} J(\theta(t))  - G(u_{i,j}(t), F^{*}; \theta(t)) \bigg{)}  }_{\textrm{Noise}},
\end{equation}
where the latter term is ``noise" in the sense that
\begin{equation}
\big\langle \nabla_{\theta} J(\theta)  - G(u_{i,j}(\cdot), F^{*}; \theta )   \big\rangle = 0. 
\end{equation}
We immediately observe from (\ref{eq:grad_flowDecomposition}) that, without the noise term, (\ref{eq:grad_flowDecomposition}) would just be deterministic gradient descent and would converge to a local minimizer of $J(\theta)$. The gradient descent algorithm (\ref{eq:grad_flowDecomposition}) updates the parameter in continuous time, which is often referred to as a ``gradient flow". The additional noise term makes (\ref{eq:grad_flowDecomposition}) a form of stochastic gradient descent in continuous-time (SGDCT)~\cite{Sirignano2017}.  

Constructing an estimator $G(u_{i,j}(t), F^\ast; \theta)$ for $\nabla_{\theta} J(\theta)$ is non-trivial due to the obstacles discussed earlier arising from the chaotic dynamics of $u(t)$. The following estimator $G_i(u_{i,j}(t), F^\ast; \theta)$ will be used for the $i^\mathrm{th}$ parameter\footnote{The formula can be applied to ODEs (instead of PDEs) by simply omitting the spatial integral.}:
\begin{equation}\label{eq:estimator_G}
    G_{i}(u_{i,j}(t),F^{*};\theta) = 
    \int_{\Omega} 
    2\left(f(u_{i,0}(x, t)) - F^{*}(x)\right)
    \frac{f(u_{i,1}(x, t)) - f(u_{i,2}(x, t))}{2\varepsilon_i}\, dx,
\end{equation}
where $u_{i,0}$, $u_{i,1}$, and $u_{i,2}$ are independent solution trajectories with independent random initial conditions. 

The parameter $\varepsilon$ is a hyperparameter. As $\varepsilon \rightarrow 0$, the estimate diverges due to the chaotic nature of the problem. Hence, it both introduces a $O(\varepsilon^{2})$ error into the estimate and also regularizes the gradient and therefore should be kept to a moderate magnitude. We use $\varepsilon/\theta^* \sim \mathcal{O}(0.1)$ (where $\theta^*$ is a rough estimate for the range of sensible $\theta$) for most of the subsequent cases.

Convergence of online gradient flows similar to \eqref{eq:grad_flow} to a local minimum in the limit $t \rightarrow \infty$ has been rigorously proved for stochastic differential equations~\cite{Wang2022,Sharrock2023,Sharrock2023a,Wang2024}, subject to the usual assumptions for the learning rate schedule $\alpha(t)$~\cite{Sirignano2022, Benveniste1990,Kushner2010} for convergence of gradient descent: 
\begin{equation} \label{LRcondition}
    \int_{0}^{\infty} \alpha(t)\ dt = \infty\ \text{ and } \int_{0}^{\infty} \alpha^{2}(t)\ dt < \infty.
\end{equation}
The focus of this paper is the development and numerical evaluation of the online gradient flow algorithm for chaotic dynamics (including turbulent flows). A rigorous mathematical convergence analysis, as in~\cite{Wang2022,Wang2024}, is planned as a topic for future research. Typical mathematical analysis requires the development and analysis of a Poisson equation for the fluctuation/noise term in the online gradient flow, which for deterministic chaotic dynamical systems is a non-trivial mathematical challenge. 

For all of the numerical cases in this paper, we use a learning rate of the following form
\begin{equation} \label{eq:LR}
    \alpha(t) = \begin{cases}
        \hfil\alpha_0 & 0 \leq t\leq t_\mathrm{decay}\\  
        \displaystyle\frac{\alpha_0}{1 + \alpha_1(t - t_\mathrm{decay})} & t > t_\mathrm{decay}
    \end{cases}
\end{equation}
with $\alpha_{0} > 0$ and $\alpha_{1} > 0$. The learning rate (\ref{eq:LR}) satisfies the condition (\ref{LRcondition}).

\subsection{Demonstration that the online estimator $G$ is unbiased}
For convergence of the OGF to a local minimum, it is essential that the gradient estimator $G$~\eqref{eq:estimator_G} is unbiased (defined in~\eqref{eq:bias}) as $t\rightarrow\infty$.
Verifying that this is the case is not trivial in chaotic systems.
To that end, the terms responsible for bias in the time-average of $G$ can be identified, and it can be seen that, for the example chaotic system we consider, they decay to zero in proportion to $t^{-1/2}$---meaning that $G$ is indeed asymptotically unbiased as $t\rightarrow\infty$.
We use the following abbreviations for notational convenience:
\begin{align*}
    f_{i,0} &= f(u_{i,0}(x, t)) \notag\\
    f_{i,1} &= f(u_{1,0}(x, t))\\
    f_{i,2} &= f(u_{i,2}(x, t)).
\end{align*}
To verify that $G$ is unbiased (i.e. \eqref{eq:bias} is satisfied), we consider the time-average of \eqref{eq:estimator_G},
\begin{equation}
    \big\langle G_{i}(u_{i,j}(\cdot), F^{*}; \theta) \big\rangle = \left\langle\int_{\Omega} 
    2\left(f_{i,0} - F^{*}(x)\right)
    \frac{f_{i,1} - f_{i,2}}{2\varepsilon_i}\, dx \right\rangle.
\end{equation}
Interchanging the time-average and the spatial integral yields
\begin{equation}
    \big\langle G_{i}(u_{i,j}(\cdot), F^{*}; \theta) \big\rangle =\int_{\Omega} 
    2  \left\langle\left(f_{i,0} - F^{*}(x)\right)
    \frac{f_{i,1} - f_{i,2}}{2\varepsilon_i} \right\rangle \, dx.
\end{equation}
By applying the Reynolds decomposition to $f_{i,0}$, $f_{i,1}$, and $f_{i,2}$, these terms can be written as the sum of their time-average and fluctuating components, for example:
\begin{align}\label{eq:Reynolds}
f_{i,0} &= \langle f_{i,0} \rangle  + f_{i,0}^\prime, \notag \\
f_{i,0}^\prime &= f_{i,0} - \langle f_{i,0} \rangle,  \\
\langle f_{i,0}^\prime \rangle &= 0.\notag
\end{align}
Inserting the Reynolds decomposition into the expression for the time-average gradient estimator yields
\begin{equation}\label{eq:bias_terms}
    \big\langle G_{i}(u_{i,j}(\cdot), F^{*}; \theta) \big\rangle = 
    \underbrace{\int_\Omega 2 \left(\big\langle f_{i,0}\big\rangle - F^{*} \right) \frac{\big\langle f_{i,1}\big\rangle - \big\langle f_{i,2}\big\rangle}{2\varepsilon_i}\,dx }_{\nabla_{\theta} J(\theta) + \mathcal{O}(\epsilon^2)}
    +\int_\Omega \frac{\big\langle f_{i,0}^\prime f_{i,1}^\prime \big\rangle - \big\langle f_{i,0}^\prime f_{i,2}^\prime\big\rangle}{\varepsilon_i} \,dx.
\end{equation}
The first term is $\nabla_{\theta} J(\theta) + \mathcal{O}(\varepsilon^2)$, where $\varepsilon = \max_i \varepsilon_i$. To recover an $\varepsilon^2$-accurate approximation
for \eqref{eq:bias}, the second term has to be equal to zero. In chaotic systems, we expect the correlations $\big\langle f^\prime_{i,0}f^\prime_{i,1}\big\rangle$ and $\big\langle f^\prime_{i,0}f^\prime_{i,2}\big\rangle$ to both be zero in the limit $t\rightarrow\infty$ due to Lyapunov divergence, satisfying this condition.
This should hold in systems with some coherent component (e.g. periodic forcing or cylinder vortex shedding) given initial conditions randomly selected from the initial ergodic state and a sufficiently large minibatch.

\begin{figure}[t]
    \centering
    \includegraphics[width=0.9\linewidth]{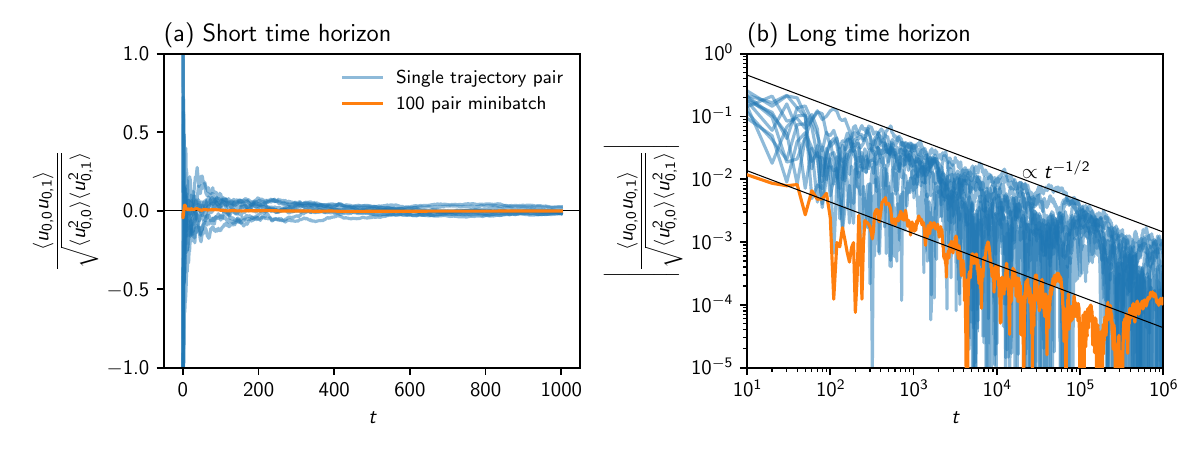}
    \vspace{-11pt}
    \caption{Decay of the normalized correlation $\left.\big\langle u^\prime_{i,0}u^\prime_{i,1}\big\rangle \middle/ \sqrt{\langle u^{\prime 2}_{i,0}\rangle\langle u^{\prime 2}_{i,1}\rangle} \right.$ (see \eqref{eq:bias_terms}) in the Lorenz-63 system \eqref{eq:lorenz63}. Black lines in (b) indicate a decrease in $\left|\big\langle u^\prime_{i,0}u^\prime_{i,1}\big\rangle\right|$ that is proportional to $t^{-1/2}$.}
    \label{fig:lorenz_decorrelation}
\end{figure}

As an example, we consider the convergence of $\big\langle u^\prime_{i,0}u^\prime_{i,1}\big\rangle$ (i.e., with $f(u)= u$) to zero in the Lorenz-63 system. 
Figure~\ref{fig:lorenz_decorrelation} shows the cumulative average of $\big\langle u^\prime_{0,0}u^\prime_{0,1}\big\rangle$ for the $x$ variable in \eqref{eq:lorenz63}, with the trajectory $u^\prime_{0,0}$ corresponding to $\theta = \lbrace \rho^\ast,\sigma^\ast,\beta^\ast\rbrace^\top$, and the trajectory $u^\prime_{0,1}$ corresponding to $\theta + \varepsilon_0e_0 = \lbrace \rho^\ast+\varepsilon_0,\sigma^\ast,\beta^\ast\rbrace^\top$ (with $\varepsilon_0=1$).
Figure~\ref{fig:lorenz_decorrelation}a shows the convergence over a short time interval $[0, 10^3]$, where the minibatch average (see Section~\ref{Sec:noise}) over 100 trajectory pairs already goes to zero quickly.
Figure~\ref{fig:lorenz_decorrelation}b shows that, as expected, the absolute value of the correlation decays with $t^{-1/2}$ both with and without minibatching (although the former is roughly one order of magnitude smaller) over a longer time interval $[0, 10^6]$.
We expect these results to hold for other chaotic systems.

\subsection{Key features of the online estimator $G$}

Aside from being asymptotically unbiased, the estimator $G$ in (\ref{eq:estimator_G}) has several other key features. First, it is completely \emph{online} since, at the current time $t$, we use the current \emph{instantaneous} values of the solution $u(x,t; \theta(t))$ to estimate the direction of steepest descent---instead of taking a time-average from the current time $t$ all the way back to time zero. Furthermore, this estimate naturally evolves as the parameter $\theta$ is updated---i.e., the instantaneous solution used in $G$ continually adjusts as $\theta$ is update---whereas time-averaging back to time zero would time-average over solutions $f \big{(} u(x,\tau, \theta(\tau) ) \big{)} |_{\tau=0}^t$ corresponding to older values of the continuously optimized $\theta(\tau)$.   

In order to obtain this online estimate, we first carefully use chain rule on the original objective function to decompose the gradient into the product of two terms, an error term and an instantaneous gradient term: 
\begin{equation}    
\Gamma(x; \theta) = 2 \Big(\langle f(u(x,\cdot\,;\theta)) \rangle - F^{*}(x) \Big) \times  \frac{\partial}{\partial \theta_i} \Big\langle f(u(x,\cdot\,;\theta)) \Big\rangle.
\end{equation}
An asymptotically unbiased estimate of this product is developed using three independent realizations of the solution (i.e., different random initial conditions): 
\begin{equation}
\Lambda(x,t; \theta) = 2\left(f(u_{i,0}(x, t; \theta)) - F^{*}(x)\right) \times \frac{f(u_{i,1}(x, t; \theta + \varepsilon_i e_i)) - f(u_{i,2}(x, t; \theta - \varepsilon_i e_i))}{2\varepsilon_i}.
\end{equation}
The independent solution paths $(u_{i,0}, u_{i,1}, u_{i,2})$ are essential as then the time-average of $\Lambda(x,t; \theta)$ accurately approximates $ \Gamma(x; \theta)$ (up to some small finite difference error of $\mathcal{O}( \varepsilon^2)$). That is, the time-average of the product of instantaneous terms in $\Lambda(x,t; \theta)$ will equal the product of the time-average of the instantaneous terms. This would not be true if the same solution path was used for all the terms.

The finite-difference estimate for $ \frac{\partial}{\partial \theta_i} \Big\langle f(u(x,\cdot\,;\theta)) \Big\rangle$ allows for the interchange of the time-average and the derivative:
\begin{align}
\Big\langle \frac{f(u_{i,1}(x, \cdot; \theta + \varepsilon_i e_i)) - f(u_{i,2}(x, \cdot; \theta - \varepsilon_i e_i))}{2\varepsilon_i} \Big\rangle &= \frac{ \langle f(u_{i,1}(x, \cdot; \theta + \varepsilon_i e_i)) \rangle - \langle f(u_{i,2}(x, \cdot; \theta - \varepsilon_i e_i)) \rangle }{2\varepsilon_i} \notag \\
&= \frac{\partial}{\partial \theta_i} \langle f(u(x, \cdot; \theta )) \rangle  + \mathcal{O}(\epsilon^2),
\end{align}
whereas differentiating would not allow this interchange:
\begin{align}
\Big\langle \frac{\partial}{\partial \theta} f(u(x, \cdot; \theta)) \Big\rangle \neq  \frac{\partial}{\partial \theta} \langle f(u(x, \cdot; \theta )) \rangle. 
\end{align}

\subsection{Extensions to accelerate convergence in practice}

We implement several extensions of the OGF algorithm presented in the previous sections to accelerate convergence in practice. 

\subsubsection{Reducing the noise in $G$} \label{Sec:noise}
Reducing the noise in the gradient estimator $G$ allows for a larger learning rate to be used, which in turn leads to faster convergence. We address this by either using a minibatch or an exponential-weighted moving average (EWMA). EWMAs are an attractive approach for averaging as they do not require a solution history to be stored, which can have a large memory cost. In practice, the combination of a medium-size minibatch and a moderate-length moving window size can effectively reduce the noise while keeping the computational cost manageable.
As seen in Fig.~\ref{fig:lorenz_decorrelation}b, minibatching is very effective at reducing the size of the correlation term in~\eqref{eq:bias_terms}. We explore the effect of different minibatch sizes and EWMA lengths on the numerical performance of the algorithm in Section~\ref{sec:hyperparameter_sensitivity}. 

For a minibatch of size $N_{mb}$, we estimate the loss gradient at time $t$ by
\begin{equation} \label{eq:lossGradEst}
    G_{i}(u_{ij}, F^{*};\ \theta)(t) = \int_\Omega
    \Bigg[\frac{1}{N_{mb}} \sum_{n=1}^{N_{mb}} A_{n,i}(x,\ t)\Bigg] 
    \Bigg[\frac{1}{N_{mb}} \sum_{n=1}^{N_{mb}} B_{n,i}(x,\ t)\Bigg]\,dx,
\end{equation}
where $A_{n,i}(x,\ t)$ and $B_{n,i}(x,\ t)$ are the EWMAs of the first and second terms in \eqref{eq:estimator_G} for minibatch sample $n$. They are updated at each time step according to
\begin{equation}\label{eq:ewma}
\begin{aligned}
    A_{n,i}(x,\ t) &= \frac{2}{M+1}\Big(f(u_{i,0}(x,\ t)) - F^{*} \Big) + \frac{M}{M+1}A_{n,i}(x,\ t - \Delta t)  \\
    B_{n,i}(x,\ t) &= \frac{1}{M+1}\left(\frac{f(u_{i,1}(x,\ t)) - f(u_{i,2}(x,\ t))}{2\varepsilon_i}\right) + \frac{M}{M+1}B_{n,i}(x,\ t - \Delta t),
\end{aligned}
\end{equation}
where $M$ is the nominal EWMA length in terms of number of time steps\footnote{To be time step independent, this can easily be reformulated into uncoupled ODEs for $A_{n,i}$ and $B_{n,i}$.
Given a generic variable q, its EWMA $\ol{q}$ with weight $\omega$ obeys
\begin{equation*}
    \frac{d\ol{q}}{dt} = \frac{1}{\omega} \Big(q(t) - \ol{q}(t)\Big).
\end{equation*}
Applying the backward Euler time integration method, we obtain equations of the form \eqref{eq:ewma} with $M = \omega / \Delta t$.
} and $t-\Delta t$ is the time at the previous time step.

The EWMA method developed above has a key difference in comparison to the momentum mechanism used in existing gradient descent algorithms such as RMSprop~\cite{Rumelhart1986} and ADAM~\cite{Kingma2014}. Typical momentum methods would take an EWMA of the gradient estimator $G_{i}(u, F^{*};\ \theta)$ rather than applying EWMA to the individual terms in a decomposition of the gradient estimate as proposed above. We have observed that applying ``classic" momentum methods increase the noise in the gradient estimator from transient fluctuations of the last term in \eqref{eq:bias_terms} compared to the EWMA-decomposition approach, which directly estimates only the first term in \eqref{eq:bias_terms} (i.e. the term of interest for our optimization). 

To see why this EWMA-decomposition approach is beneficial, it is useful to consider a comparison to a gradient flow for a standard machine learning model with many parameters (e.g., a neural network).
In this case, the gradients of the parameters are typically computed based upon the same data sample. In contrast, the OGF gradient estimates are inherently independent, since the trajectories used in \eqref{eq:estimator_G} are independent. That is, each gradient calculation uses a different independent data sample. For a scalar model $g(x;\theta)$, the model output changes over time according to
\begin{equation}
    \frac{\partial g}{\partial t} = \frac{\partial g}{\partial \theta} \frac{\partial \theta}{\partial t}.
\end{equation}
 In general, the variance over a minibatch is given by
\begin{equation}
    \text{Var}\left(\frac{\partial g}{\partial t}\right) = \sum_{i} \left(\frac{\partial g}{\partial \theta}\right)^{2}_{i} \Sigma_{ii} +2\sum_{i<j} \left(\frac{\partial g}{\partial \theta}\right)_{i} \left(\frac{\partial g}{\partial \theta}\right)_{j} \Sigma_{ij},
\end{equation}
where $\Sigma$ denotes the covariance matrix of $\frac{\partial \theta}{\partial t}$. Due to the independence of gradient estimates, the second sum vanishes for our method. We conjecture that this leads to a smaller variance of the trained model outputs.

\subsubsection{Adaptive optimization algorithms}
In many situations (particularly with multiple parameters that may have very different gradient magnitudes) it is desirable to use an adaptive optimization algorithm such as RMS\-prop~\cite{Hinton2012}.
This can be easily included within the present framework by advancing the parameters in time using
\begin{equation}
    \dd{\theta}{t} = \frac{- \alpha(t)}{\sqrt{\widetilde{G^2}(t) + \epsilon}}\odot G(u_{ij}, F^\ast; \theta),
\end{equation}
where $\odot$ is the elementwise product, $\epsilon = 1\times 10^{-8}$ is used to prevent division by zero and the moving average of the squared gradient is updated with an EWMA according to
\begin{equation}\label{eq:rmsprop_g2}
    \widetilde{G^2}(t) = (1-\beta_1\ )G(t)^2 + \beta_1 \widetilde{G^2}(t - \Delta t).
\end{equation}
We typically use $\beta_1 = 0.99$.

\subsection{Discussion of the methodology}
The primary advantages of the online gradient flow (OGF) method are:
\begin{itemize}
    \item The algorithm scales well to systems with a large number of degrees of freedom. This is because it uses standard time-marching forward algorithms that do not suffer from the curse of dimensionality. Therefore, it does not require the solution of large systems of KKT equations over the space-time domain. Due to its online nature, it does not require iterating over many time-averaged realizations of the flow and is able to update the gradient estimate as fast as the system reacts to changes in the parameters.
    \item It permits arbitrary time horizons. This is necessary for chaotic systems with a broad spectrum of time scales. Standard offline optimization methods (including adjoint, stabilized adjoint, and offline finite difference methods) for optimizing turbulent flows/chaotic dynamics (see \citet{LEA2000}, \citet{Liu2023}, \citet{Liu-2025}, and \citet{Garai2021}) require an optimization horizon $\tau_{opt}$ to be specified \emph{a priori} --- this must be long enough to be representative of the gradient in the limit $t\rightarrow\infty$, but short enough for many iterations to be computationally feasible. A suitable minimum value for $\tau_{opt}$ is case dependent and requires an expensive process of trial and error to identify. Furthermore, the computational cost for long time horizons necessary to represent the time scales can be computationally intractable to simulate for many optimization iterations. In contrast, the OGF method directly optimizes over the infinite time interval $[0, \infty)$ using an online estimator for the direction of steepest descent \emph{in a single simulation} without requiring many sequential optimization iterations over long time-length simulations.
    \item In contrast to some approaches that work with approximate periodic solutions or reduced-order models of the system, we solve the governing equations exactly.
    \item It is easy to implement in existing flow solvers. It does not require adjoint solutions, and the only communication between solver trajectories is at the gradient calculation/parameter update stage, which could feasibly be moved to a top-level driver script outside of the flow solver code.
    \item Due to being an online algorithm, it does not require additional storage of the solution history. Adjoint-based approaches must store the whole solution trajectory to backpropagate over, which can require storing petabytes of data for long optimization time horizons. While checkpointing strategies exist that alleviate this somewhat~\cite{Griewank2008}, these add further complexity and computational cost to the code.
\end{itemize}

The predominant limitation of the algorithm is its linear cost scaling with the number of parameters (in contrast to methods focusing on stabilizing the adjoint).
This complicates optimizing over large neural networks, though we note that the algorithm is perfectly parallel (in terms of the number of parameters), so large numbers of parameters would be feasible with sufficient parallel computing resources.
The multiple solution trajectories required for the gradient estimator and minibatching ($3 \times N_{mb}$ in total) can be further parallelized effectively by GPUs. We exploit this for the large-minibatch computations of the Kuramoto--Sivashinsky PDE in Section~\ref{sec:kse} and forced homogeneous isotropic turbulence in Section~\ref{s4}.

In the rest of this paper, we illustrate the effectiveness of our algorithm in applications of increasing complexity and size.
Sections~\ref{s3}~and~\ref{sec:kse} cover low-dimensional systems commonly studied in the literature. In Section~\ref{s4}, we present applications to large degree of freedom turbulent flow solutions.

\section{Lorenz-63 system of ODEs}\label{s3}\label{sec:lorenz}
\subsection{Case description}
For our first test case, we consider the Lorenz-63 system system of ODEs derived in~\cite{Lorenz1963} (and stated in this paper in equation \eqref{eq:lorenz63}) as a reduced-order model for natural convection.
This is a very popular first application for optimization algorithms in chaotic systems, presumably due to its small number of degrees of freedom $u(t) = \lbrace x(t),\ y(t),\ z(t) \rbrace^\top$ and parameters $\theta = \lbrace \rho,\ \sigma,\ \beta \rbrace^\top$.

For testing the OGF methodology, it is useful to have a known value for the optimal parameters.
We therefore select the target parameters as the classical values from Lorenz's original paper~\cite{Lorenz1963}, i.e. $\rho^{*} = 28$, $\sigma^{*} = 10$, and $\beta^{*} = 8/3$.

We wish to select a parameter $\theta$ to minimize the objective function
\begin{equation}\label{eq:lorenz_J}
    J(\theta) = 
    \left(\frac{\langle x^2 \rangle - \langle x^2\rangle^{*}}{\langle x^2\rangle^{*}}\right)^2 + 
    \left(\frac{\langle y^2 \rangle - \langle y^2\rangle^{*}}{\langle y^2\rangle^{*}}\right)^2 +
    \left(\frac{\langle z^2 \rangle - \langle z^2\rangle^{*}}{\langle z^2\rangle^{*}}\right)^2,
\end{equation}
where $\langle x^2\rangle^{*} = 67.84$, $\langle y^2\rangle^{*} = 84.02$, and $\langle z^2\rangle^{*} = 689.0$ are the time-averaged values of $x(t)^2$, $y(t)^2$, and $z(t)^2$ with the target parameter values $\rho^\ast$, $\sigma^\ast$, and $\beta^\ast$.
The time average of $x^2$ is used as the target statistic for our optimization (rather than $x$) because the time-averaged values of $x$ and $y$ are zero for all parameter values, and fixing $\langle x^2 \rangle$ is sufficient to fix $\langle z \rangle$ (given fixed parameters)~\cite{Lucke1976}. 
The target statistics were calculated over a very large time horizon of 1,000 time units after reaching the ergodic state and a very large minibatch of 100,000 randomly initialized trajectories to eliminate statistical error.

Unless otherwise specified, we use finite difference perturbation sizes of $\varepsilon = 1$ for $\rho$ and $\sigma$, and $\varepsilon = 0.1$ for $\beta$, $t_{decay} = 200$, and $\alpha_1$ is set so that the learning rate is decayed from its initial value of $\alpha_0 = 0.1$ to 0.01 over 1,000 time units (see~\eqref{eq:LR}).  This value of $t_{decay}$ was chosen to be sometime after the optimization had converged to an neighborhood around the target parameter values, demonstrating the stability of our method with a fixed learning rate.

\subsection{Optimization results}
\begin{figure}[t]
    \centering
    \includegraphics[width=0.9\linewidth]{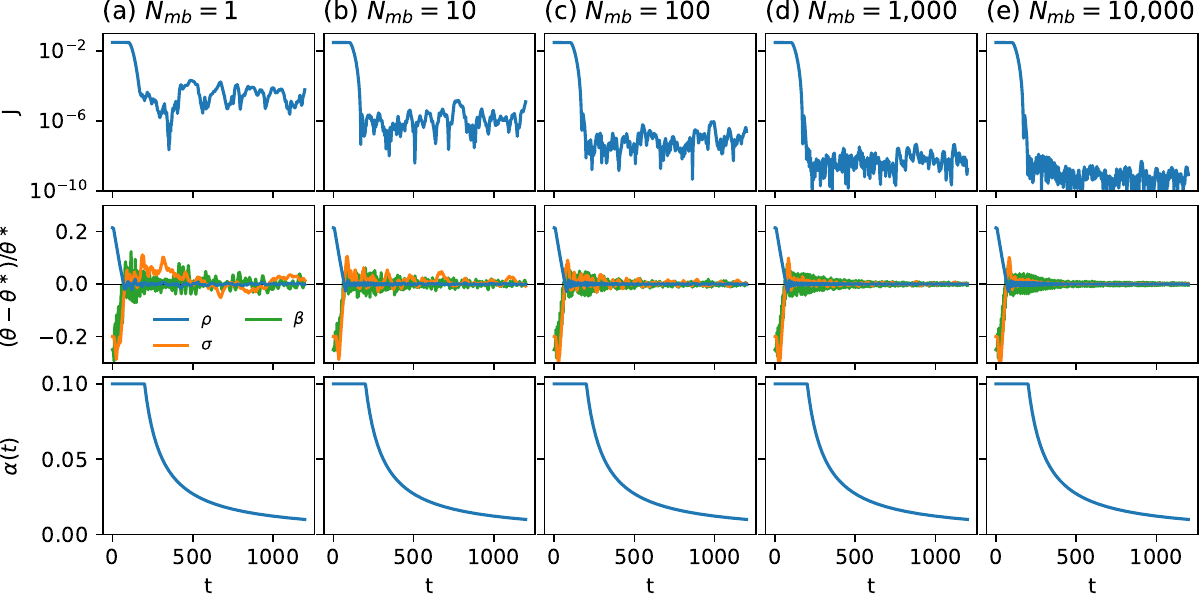}
    \caption{Optimization of the Lorenz-63 system parameters using the online optimization algorithm for a range of different minibatch sizes $N_{mb}$ across 4 orders of magnitude. Row one displays the 200 time unit loss estimate $J$, row two reports the relative parameter error $(\theta - \theta^\ast)/\theta^\ast$, and row three shows learning rate schedule $\alpha(t)$. All cases use $M = 1,000$.}
    \label{fig:lorenz_general_plot}
\end{figure}

Figure~\ref{fig:lorenz_general_plot} displays the convergence of the online gradient descent methodology for a range of different minibatch sizes.
The objective function $J$ is estimated at each time instant using a 200 time unit moving average of the solution.
The methodology robustly converges for all minibatch sizes, decreasing the objective function by several orders of magnitude, and converges the parameters to a precision that is more than acceptable for any foreseeable application of the method to closure modeling or control (even without the use of minibatching). As might be expected there is a trade-off to be made in terms of expensive but low-noise convergence to the target parameters with large minibatch sizes (see e.g. Fig.~\ref{fig:lorenz_general_plot}e) compared to a noisy but computationally cheap final estimate of the optimal parameters with smaller minibatch sizes as in Fig.~\ref{fig:lorenz_general_plot}a.

Table~\ref{tab:lorenz_param_errs} uses the normalized root mean squared error (RMSE) of the parameters over the last 100 time units, defined for a parameter $\theta$ and length $\tau$ as
\begin{equation} \label{eq:RMSE}
    \text{RMSE}^*(\theta) = \frac{1}{\theta^{*}}\sqrt{\int_{t_{end}-\tau}^{t_{end}} (\theta(t) - \theta^{*})^2\ dt}.
\end{equation}
This is a good performance metric for the optimization algorithm as it penalizes both differences in the average value of the parameter and fluctuations of the parameter about its target value.
An estimate for the loss function, again calculated using a time-average of the solution over the last 100 time-units, is given in Table~\ref{tab:lorenz_J}.
As shown in Tables~\ref{tab:lorenz_param_errs}~and~\ref{tab:lorenz_J}, the target parameters are recovered to within an error of 1\% with an appropriate combination of $N_{mb}$ and $M$. The loss is also reduced by multiple orders of magnitude.
\begin{table}[t]
    \centering
    \caption{Normalized RMSE averaged across all 3 Lorenz-63 parameters over the last 100 time-units of the optimization for different minibatch sizes $N_{mb}$ and EWMA lengths $M$.}\label{tab:lorenz_param_errs}
    \begin{tabular}{rr| llll}
        \toprule
        && \multicolumn{4}{c}{$M$} \\
        && 1 & 10 & 100 & 1,000\\
        \midrule
        \multirow{5}*{$N_{mb}$}& 1    & 19.9\% & 16.7\% & 1.73\% & 0.809\% \\ 
        & 10   & 6.50\% & 3.53\% & 0.341\% & 0.629\% \\
        & 100  & 0.637\% & 1.20\% & 0.235\% & 0.432\% \\
        & 1,000 & 0.970\% & 0.837\% & 0.103\% & 0.219\% \\
        & 10,000 & 0.551\% & 0.449\% & 0.0831\% & 0.148\% \\
        \bottomrule
    \end{tabular}
\end{table}
\begin{table}[t]
    \centering
    \caption{Loss function $J$ estimated over the last 100 time-units of the optimization for different minibatch sizes $N_{mb}$ and EWMA lengths $M$.}\label{tab:lorenz_J}
    \begin{tabular}{rr| llll}
        \toprule
        && \multicolumn{4}{c}{$M$} \\
        && 1 & 10 & 100 & 1,000\\
        \midrule
        \multirow{5}*{$N_{mb}$}& 1 &  $2.57\times10^{-1}$ & $5.22\times10^{-1}$ & $1.26\times10^{-3}$ & $7.12\times10^{-6}$ \\
        & 10 & $3.17\times10^{-2}$ & $9.80\times10^{-3}$ & $3.37\times10^{-5}$ & $5.57\times10^{-6}$ \\
        & 100 & $8.19\times10^{-5}$ & $3.97\times10^{-4}$ & $7.56\times10^{-7}$ & $1.99\times10^{-7}$ \\
        & 1,000 & $6.78\times10^{-5}$ & $5.02\times10^{-5}$ & $7.93\times10^{-7}$ & $1.31\times10^{-9}$ \\
        & 10,000 & $1.25\times10^{-5}$ & $1.29\times10^{-5}$ & $2.20\times10^{-8}$ & $5.15\times10^{-10}$ \\
        \bottomrule
    \end{tabular}
\end{table}

The results in this section constitute a first demonstration of the OGF methodology to the Lorenz-63 system, a relatively simple chaotic ODE. We have shown that the method is been able to accurately converge to the \emph{a priori} known optimal parameters of the system as long as sufficient noise reduction (in terms of the combined effects of the EWMA and minibatching) is used. We next apply the OGF method to a single-variable PDE,  the modified Kuramoto--Sivashinsky equation. This has more degrees of freedom than the Lorenz-63 ODE system (513 mesh points versus 3), and being a PDE requires the inclusion of the spatial integral in the gradient estimator.

\section{Kuramoto--Sivashinsky PDE}\label{sec:kse}
\subsection{Governing equations and numerical solver}
In this section, we apply the online optimization method to the modified Kuramoto--Sivashinsky equation (KSE). The KSE was derived independently by Kuramoto \cite{Kuramoto-1976,Kuramoto1978} as a model for angular-phase turbulence in the context of a 3D reaction-diffusion equation and by Sivashinsky \cite{Sivashinsky1977,Michelson-1977} to describe the evolution of diffusive-thermal instabilities in a planar flame front.
The modified KSE we consider is a fourth-order, chaotic PDE,
\begin{equation} \label{eq:KSE}
\begin{aligned}
\frac{\partial u}{\partial t} = - \Big( \nu \frac{\partial^{4}u}{\partial x^{4}}  + \frac{\partial^{2} u}{\partial x^{2}} + (u+c) \frac{\partial u}{\partial x}  \Big), \quad &\text{for} \; x \in [0,L], \; t \geq 0 \\
u(0,t) = u(L,t) = 0,\quad &\text{for} \; t \geq 0 \\
\frac{\partial u}{\partial x}(0,t) = \frac{\partial u}{\partial x}(L,t) = 0,\quad &\text{for} \; t \geq 0 \\
u(x,0) = u_{0}(x),  \quad &\text{for} \; x \in [0,L].
\end{aligned}
\end{equation}
The homogeneous Dirichlet and Neumann conditions in \eqref{eq:KSE}, which are not included in the classical KSE system, make the KSE ergodic~\cite{Blonigan2014}.

The parameter $c$ influences the convection speed of the system and has been used by~\citet{Blonigan2014}. They presented a classification of the qualitative behavior based on the values of $c$. For our purposes, it suffices to restrict ourselves to $c$ between $0$ and $1.2$. \citet{Blonigan2014} found the solution to exhibit chaotic behavior in this regime.
We also introduce the diffusion parameter $\nu$ as done in \cite{Hyman1986}. 
The interval length is fixed at $L=128$ to ensure chaoticity of the system for $\nu$ larger than $0$ and less than $2$~\cite{Hyman1986}, which is sufficient for our purposes. 
We optimize over the parameters $c$ and $\nu$ and keep $L$ fixed.

\begin{figure}[t]
    \centering
    \includegraphics[width=0.75\linewidth,trim={0 20 0 0},clip]{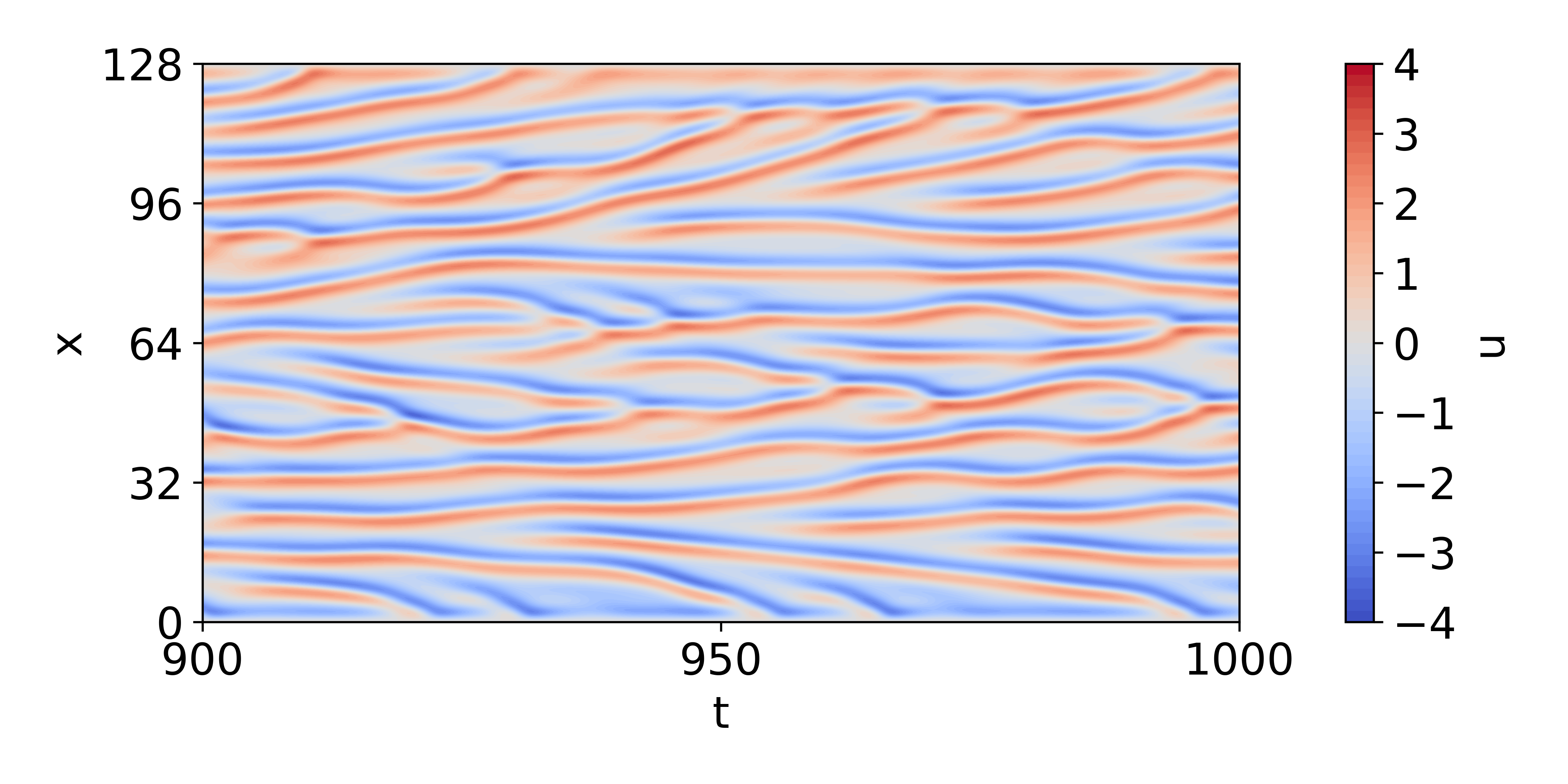}
    \caption{KSE solution $u$ with $c=0.2$ and $\nu = 1.0$ for $t \in [900,1000]$. }
    \label{fig:KSE_fwd}
\end{figure}

The spatial interval $[0,L]$ is discretized uniformly by a set of $N_{x} = 513$ points.
Spatial derivatives are discretized using second-order central differences, and time integration is performed using the implicit-explicit Runge-Kutta scheme IMEXRK34S[2R]L$\alpha$ from \cite{Cavaglieri-2013}, with $\Delta t = 0.1$. Figure~\ref{fig:KSE_fwd} shows a single solution trajectory for the KSE.

\subsection{Optimization results}
\begin{figure}[t]
    \centering
    \includegraphics[width=0.6\linewidth]{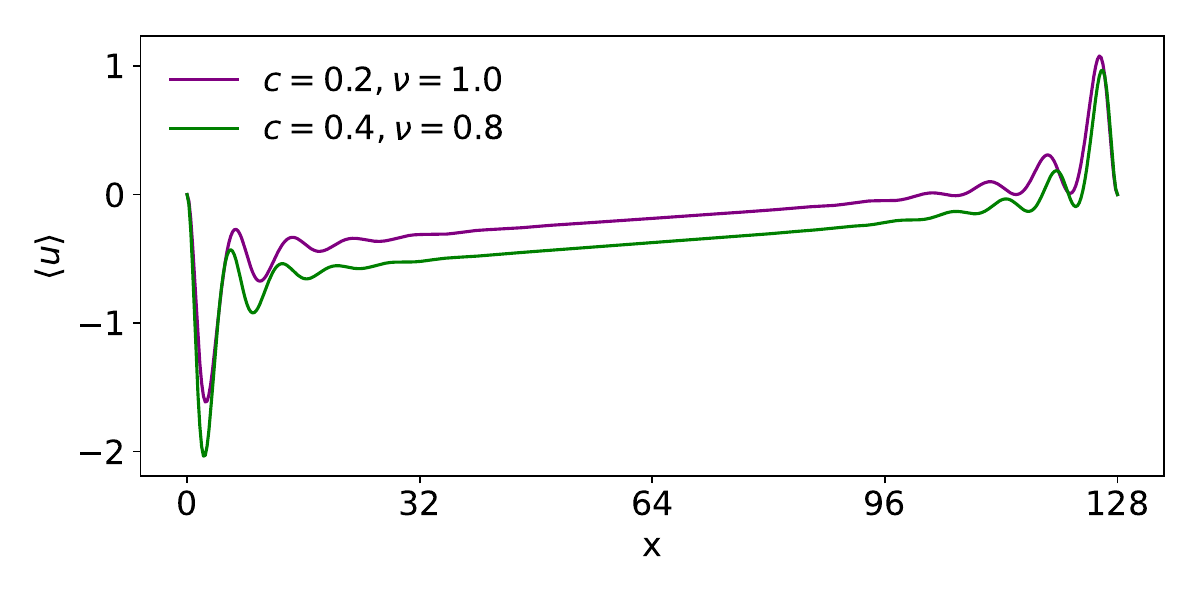}
    \caption{The mean profile $\langle u \rangle $, averaged over $11,000$ time units and $10,000$ minibatch samples. }
    \label{fig:KSE_ergodics}
\end{figure}

We optimize over the parameters $c$ and $\nu$ of the KSE to minimize the squared $L^2$ loss
\begin{equation}
J(\theta) = \int_{\Omega} \Big( \langle u(x,\cdot;\theta) \rangle -  u^{*}(x)\Big)^{2} dx.
\end{equation}
In the optimization experiment we present, the target data is the time and ensemble average of a KSE solution $u^{*}$, generated with $c^{*} = 0.4$ and $\nu^{*} = 0.8$. The field $u^{*}$ is averaged over the interval from $t_{0}=1,000$ to $t_{1} = 12,000$ and over $N_{mb} = 10,000$ samples. The average profile of $u$ with the initial parameters and the profile of $u^*$ can be seen in Fig. \ref{fig:KSE_ergodics}.

The objective is to recover the target parameters $(c^{*}, \nu^{*})$ from our simulation with initial parameters $(c, \nu) = (0.2,1.0)$.  The effect of different choices for $N_{mb}$ and $M$ will be examined in Section \ref{sec:hyperparameter_sensitivity}.
As for the Lorenz system, we evaluate the performance using the RMSE \eqref{eq:RMSE} for measuring the parameters' convergence and provide an estimate for the loss function $J$ over a small time interval of $1,000$ time-units.

We use the RMSProp optimizer with $\beta_{1} = 0.99$, $\varepsilon = 0.1$, the learning-rate schedule \eqref{eq:LR} with
\begin{equation}
    \alpha_{0} = 5\times10^{-5}, \quad \alpha_{1} = 6\times10^{-4}, \quad t_{decay} = 5,000,
\end{equation}
selected minibatch sizes $N_{mb}$ between $1$ and $1,000$, and EWMA hyperparameters $M$ between $1$ and $5,000$.

 The value for the hyperparameter $\alpha_{0}$ is the inverse of twice the number of iterations it took our numerical solver to reach the ergodic state. 
 The factor of two is added as a safety margin.
 The value $\alpha_{1}$ is chosen such that $\alpha(t)$ is approximately one tenth of the initial learning rate after 15,000 time units: $\alpha(t_{decay} + 15,000) \approx \alpha_{0} / 10$. We have found this set of hyperparameters to be generally effective for the KSE at reducing oscillations of the optimized parameters around their respective target values during training. We found the algorithm to exhibit some robustness to the finite difference step size $\varepsilon$ and learning rate $\alpha$, which has also been examined in Sec. \ref{s3}. Hence, we focus on evaluating the performance for different choices of the minibatch size $N_{mb}$ and EWMA length $M$.

\begin{figure}[t]
    \centering
    \includegraphics[width=0.9\linewidth]{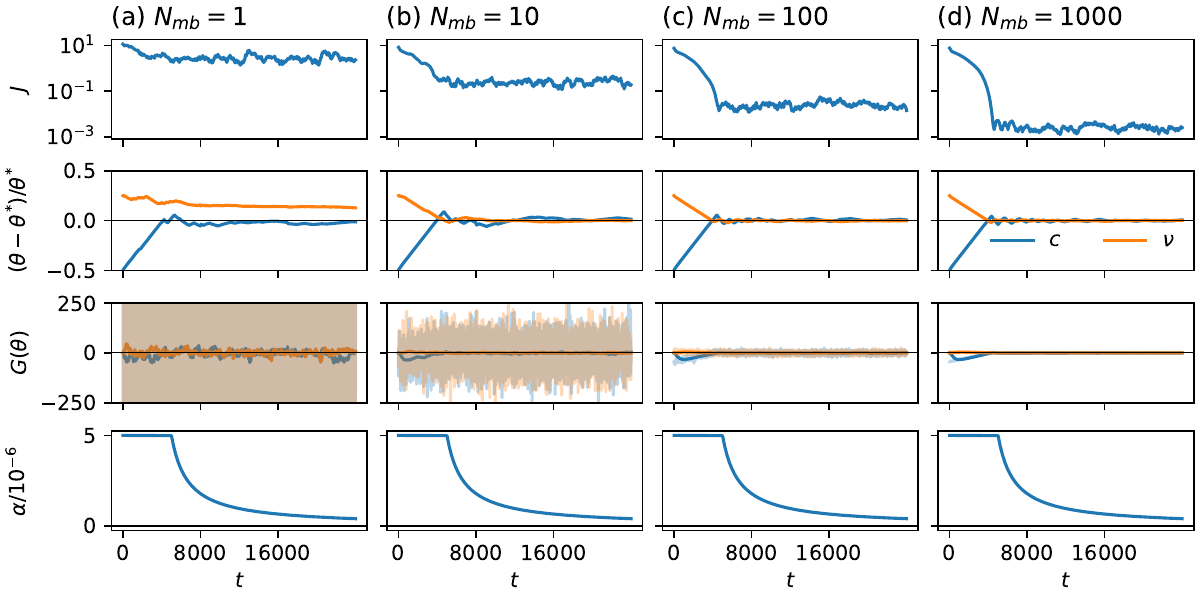}
    \caption{KSE loss function $J(\theta)$ estimated over $1,000$ time units, optimization parameter error $(\theta - \theta^{*})/\theta^{*}$, gradient estimate $G$, and learning rate $\alpha(t)$ for $M=3,000$. Saturated lines correspond to the EWMA estimates, light lines are instantaneous values. } 
    \label{fig:KSE_comp_mb}
\end{figure}

We include plots of the loss estimate, parameter evolution, gradient estimates (instantaneous and averaged), and the learning rate for different minibatch sizes $N_{mb}$ in Fig.~\ref{fig:KSE_comp_mb}. 
The algorithm converges to a small error ball around the target parameter within approximately 5,000 time units, except for $\nu$ with $N_{mb}=1$. An increase in the minibatch size reduces the noise in the instantaneous gradient estimate and leads to smoother convergence. The moving average gradient estimate is less affected---the biggest difference is seen between $N_{mb} = 1$ and $N_{mb} = 10$. Visually, the algorithm successfully optimizes over the parameters for $N_{mb} = 10$ and larger minibatch sizes. 

An overview of achieved loss, the RMSEs for $c$, and the RMSEs for $\nu$, for all combinations of $N_{mb}$ and $M$, are provided in Tabs.~\ref{tab:KSE_J}, \ref{tab:KSE_c_err}, and \ref{tab:KSE_nu_err}, respectively.
While Tab.~\ref{tab:KSE_c_err} shows that for $N_{mb}=1$ and $M = 1000$, $3000$, and $5000$, we are able to achieve $O(1) \%$ RMSE for $c$, we observe in Tab.~\ref{tab:KSE_nu_err} that this is not the case for $\nu$. This is in line with our observation in Fig.~\ref{fig:KSE_comp_mb}. This is due to the diffusion parameter $\nu$ having a longer characteristic time scale than $c$ and is a cautionary example: a set of hyperparameters that work well to optimize one parameter might not work to optimize another. Table \ref{tab:KSE_nu_err} shows that, for a given learning rate, it is possible to achieve a better result by increasing the minibatch size to $N_{mb}=10$. Alternatively, one can use a smaller learning rate and optimize over a longer time horizon.

\begin{table}[t]
\centering
    \caption{Loss function $J$ for the KSE estimated over the last $1,000$ time-units of the  optimization.}\label{tab:KSE_J}
    \begin{tabular}{rr| llllll}
        \toprule
        && \multicolumn{6}{c}{$M$} \\
        && 1 & 10 & 100 & 1,000 & 3,000 & 5,000\\
        \midrule
        \multirow{4}*{$N_{mb}$}
& 1 & $6.44 \times 10^0$ & $5.49 \times 10^0$ & $6.91 \times 10^0$ & $2.59 \times 10^0$ & $2.28 \times 10^0$ & $2.50 \times 10^0$  \\
& 10& $8.07 \times 10^{-1}$ & $7.30 \times 10^{-1}$ & $3.95 \times 10^{-1}$ & $2.06 \times 10^{-1}$ & $1.89 \times 10^{-1}$ & $2.50 \times 10^{-1}$ \\
& 100& $3.90 \times 10^{-2}$ & $1.80 \times 10^{-2}$ & $1.40 \times 10^{-2}$ & $1.80 \times 10^{-2}$ & $1.40 \times 10^{-2}$ & $3.10 \times 10^{-2}$  \\
& 1,000 & $2.00 \times 10^{-3}$ & $2.00 \times 10^{-3}$ & $3.00 \times 10^{-3}$ & $2.00 \times 10^{-3}$ & $2.00 \times 10^{-3}$ & $3.00 \times 10^{-3}$  \\
        \bottomrule
    \end{tabular}
\end{table}
\begin{table}[t]
    \centering
    \caption{RMSE in percent for the KSE of the convection parameter $c$ computed over the last $1,000$ time-units of the optimization.}\label{tab:KSE_c_err}
    \begin{tabular}{rr| llllll}
        \toprule
        && \multicolumn{6}{c}{$M$} \\
        && 1 & 10 & 100 & 1,000 & 3,000 & 5,000\\
        \midrule
        \multirow{4}*{$N_{mb}$}
        &1& 40.01\% & 43.01\% & 32.67\% & 0.48\% & 1.18\% & 0.88\%  \\
&10& 7.43\% & 8.63\% & 1.44\% & 0.41\% & 1.91\% & 0.34\%  \\
&100& 0.62\% & 0.14\% & 0.91\% & 0.09\% & 1.13\% & 0.41\%  \\
&1,000& 0.07\% & 0.18\% & 0.05\% & 0.00\% & 0.32\% & 0.57\%  \\
        \bottomrule
    \end{tabular}\\
    
    \vspace{11pt}
    \centering
    \caption{RMSE in percent for the KSE  of the diffusion parameter $\nu$ computed over the last $1,000$ time-units of the optimization.}\label{tab:KSE_nu_err}
    \begin{tabular}{rr| llllll}
        \toprule
        && \multicolumn{6}{c}{$M$} \\
        && 1 & 10 & 100 & 1,000 & 3,000 & 5,000\\
        \midrule
        \multirow{4}*{$N_{mb}$}
        &1& 24.64\% & 24.69\% & 21.29\% & 22.14\% & 13.04\% & 19.80\%  \\
&10& 18.61\% & 20.40\% & 15.93\% & 0.37\% & 0.04\% & 1.06\%  \\
&100& 2.59\% & 1.93\% & 0.15\% & 0.62\% & 0.15\% & 0.02\%  \\
&1,000& 0.05\% & 0.01\% & 0.06\% & 0.15\% & 0.22\% & 0.19\%  \\
        \bottomrule
    \end{tabular}
\end{table}

The preceding section has presented an application of the OGF method to an optimization problem governed by a low-dimensional PDE, the KSE, with greater complexity than the Lorenz-63 system. We have  discussed the learning rate selection and observe that the KSE timescales are much longer than those of the Lorenz-63 system. We also provide comparisons of the algorithm's performance for different minibatch sizes and EWMA parameters, including a cautionary example showing suboptimal performance due to insufficient averaging of the gradient estimates. 

The following section presents an application of the OGF method to forced homogeneous isotropic turbulence, a still higher-dimensional system governed by the Navier--Stokes equations, to illustrate the applicability of the method to actively studied turbulent flow problems.
\section{Forced homogeneous isotropic turbulence}\label{s4}
We now consider an application to compressible forced homogeneous isotropic turbulence (fHIT).
Compressible fHIT is a relatively complex case, for energy can be stored in acoustic or kinetic modes (our loss function is based only on the latter) before being dissipated. In addition, all of these modes are coupled~\cite{Kida1990,Jagannathan2016}. These effects are not present in incompressible flows. Compressible fHIT therefore represents a computationally and physically challenging case for optimization methods; significantly more so than the Lorenz-63 and KSE systems.

\subsection{Governing equations and solver}\label{s:GEQ}
Being a turbulent fluid flow, fHIT is governed by the three-dimensional compressible Navier--Stokes equations,
\begin{equation} \label{eq:NSE}
    \frac{\partial \bQ}{\partial t} + \nabla \cdot \bF_{c} - \nabla \cdot \bF_{v}  + \bS = 0,
\end{equation} 
where $\bS$ is a source term. The conservative variables $\bQ$, the convective flux $\bF_c$, and the viscous flux $\bF_v$ are given by
\begin{equation}
    \bQ = \begin{Bmatrix}
        \rho\\
        \rho\bu\\
        \rho E
    \end{Bmatrix},
    \quad
    \bF_{c} = \begin{Bmatrix}
        \rho \bu \\
        \rho \bu \otimes\bu + \frac{p}{\mathrm{Ma}^{2}} \bI \\
        \rho \bu \left(E + \frac{p}{\mathrm{Ma}^{2}\rho}  \right)
    \end{Bmatrix}
    \quad \text{and} \quad \bF_{v} = \frac{1}{\mathrm{Re}} \begin{Bmatrix}
        0 \\
        \sigma \\
        \bu^{\top} \sigma - \frac{1}{\mathrm{Ma}^{2} \mathrm{Pr}}  \bq
    \end{Bmatrix},
    \label{eq:NS_RHS}
\end{equation}
where $\bu = \lbrace u ,v, w \rbrace^\top$ is the velocity vector, $\rho$ is the mass density, $E$ is the total energy, $p$ is the pressure, $\bI$ the identity matrix, $\sigma = \mu(T) (\nabla \bu + \nabla \bu^{\top} - \frac{2}{3} (\nabla \cdot \bu) \bI)$ is the viscous stress tensor, and $\bq = \mu(T) \nabla T / \mathrm{Pr}$ is the heat flux vector with a constant Prandtl number $\mathrm{Pr} = 0.7$. In \eqref{eq:NS_RHS}, $\mathrm{Re}$ and $\mathrm{Ma}$ are the scaling Reynolds and Mach numbers, values of which we provide subsequently in Sec.~\ref{sec:NS_cases}.

The Navier--Stokes equations are solved using an in-house, structured, curvilinear-mesh, finite-difference solver, \emph{PyFlowCL}. 
Derivatives are computed using fourth-order central differences, and explicit time stepping is carried out using RK4.
To remove spurious odd-even oscillations introduced by the spatial discretization, the solution is filtered with an eighth-order implicit spatial filter \cite{LELE1992} at the end of each time step.
\emph{PyFlowCL} has been specially written using the \emph{PyTorch} framework to support data-driven computational fluid dynamics methodology development~\cite{Hickling2024,Nair2023,Kakka2025,Liu-2025}.
It supports full GPU acceleration, domain decomposition across multiple GPUs, minibatching (both local to a GPU and across many GPUs), and automatic differentiation for adjoint optimization.
Our proposed methodology requires very little solver modification, only needing communication between solution trajectories, and so is nonintrusive to implement in existing flow solvers such as \emph{PyFlowCL}.

\subsection{Case description} \label{sec:NS_cases}
As a first challenge for optimizing turbulent flows, we consider direct numerical simulation (DNS) of compressible forced homogeneous isotropic turbulence in a periodic cube with side length $2\pi$.
Following \citet{Kida1990} and \citet{Jagannathan2016}, a forcing term $\bF$ is included in the Navier--Stokes equations \eqref{eq:NSE} as
\begin{equation}
    \bS = \rho
    \begin{Bmatrix}
        0 \\
        \bm F\\
        \bm u \cdot \bm F + S_e
    \end{Bmatrix}.
\end{equation}
For  compressible fHIT, the viscous dissipation leads to a buildup of internal energy.
To counteract this and make the system fully ergodic, a spatially uniform internal energy sink $S_e$ is imposed to maintain a constant volume-averaged temperature in a similar manner to~\citet{Watanabe2021}.

Following \citet{Kida1990}, we force only the largest wavenumbers in each direction with random coefficients (repeated indices here denoting summation),
\begin{equation}
    F_i(\bm x, t) = A_{ij}(t) \sin x_j + B_{ij}(t) \cos x_j,
\end{equation}
where $x_j = (x_1, x_2, x_3)$. \citet{Kida1990} choose $\bm A(t)$ and $\bm B(t)$ from a normal distribution at each time step such that their mean is 0 and their variance is
\begin{equation}\label{eq:variances}
    \mathbb{E}(A^2_{ij}) = \mathbb{E}(B^2_{ij}) = 
    \begin{cases}
        \frac{2 \psi}{3 \Delta t}&\text{ if } i=j\\
        \frac{\theta}{3\Delta t}&\text{ if } i \neq j\\
    \end{cases},
\end{equation}
where $\psi$ is the dilatational component of the forcing (i.e. $\nabla \times \bm{\psi} = 0$), and $\theta$ is the solenoidal component of the forcing (i.e. $\nabla \cdot \bm{\theta} = 0$). 
For simplicity, we focus on pure solenoidal forcing with $\psi = 0$.

Sampling $\bm A(t)$ and $\bm B(t)$ from a normal distribution at each time step is not a well-formed operation for multi-stage time-advancement schemes, such as RK4, that consider intermediate states between $t$ and $t+\Delta t$ and assume that the time derivative is smooth.
We therefore calculate $\bm A(t)$ and $\bm B(t)$ from six independent Ornstein--Uhlenbeck (OU) processes (as done by~\citet{Jagannathan2016}) so that, for example,
\begin{equation}
    dA_{01}(t) = -\frac{1}{\tau} A_{01}(t) dt + \sigma dW_t,
\end{equation}
with a correlation time $\tau = 0.1$ used in all cases presented in this paper. The standard deviation in the OU processes is
\begin{equation}
    \sigma = \sqrt{\frac{\theta}{3\tau}},
\end{equation}
where $\theta$ is the parameter to be optimized.

\begin{figure}[t!]
    \centering
    \includegraphics[width=0.8\linewidth]{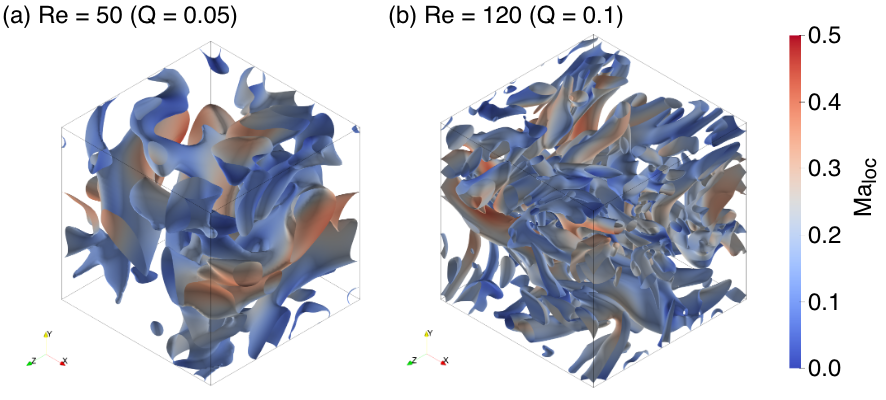}
    \caption{Isosurfaces of $Q$ (second invariant of the velocity-gradient tensor), colored by the local Mach number $\mathrm{Ma_{loc}} = |\bu| / c(T)$, for fHIT with $F_s = 1/32$ at different $\mathrm{Re}$.}
    \label{fig:fHIT_Q}
\end{figure}

We consider two cases: a low Reynolds number case, $\mathrm{Re} = 50$, and a moderately higher Reynolds number case, $\mathrm{Re} = 120$, both for scaling Mach number $\mathrm{Ma}=0.25$.
Full details of these are shown in Tab.~\ref{tab:fHIT_conditions}.
The grid size for the higher Reynolds number case corresponds to $1.7\times10^7$ grid points, or $8.4\times10^7$ degrees of freedom for the five Navier--Stokes solution variables.
The smallest Taylor Reynolds number $\mathrm{Re_\lambda}$ we consider is 48, which is sufficiently large to exhibit turbulent behavior (for instance, \citet{Overholt_1996} find turbulent behavior for $\mathrm{Re_\lambda}$ as low as 28). Visualizations of the vortical turbulent flow structures at both conditions are shown in Fig.~\ref{fig:fHIT_Q}.
Both computational grids have a grid spacing $\Delta x$ which is smaller than the Kolmogorov microscale\footnote{The Kolmogorov microscale is given by $\eta = \left({\nu^3}/{\epsilon}\right)^{0.25},$ where $\nu = \mu / \rho$ is the kinematic viscosity and $\epsilon$ is the rate of kinetic energy dissipation. As the simulation is ergodic, the rate of kinetic energy dissipation is balanced in the time-average sense by the rate of kinetic energy addition by the forcing. This is straightforwards to calculate from the standard deviation of the OU process. See \citet{Kida1990} for an example of this with random non-OU forcing.} $\eta$ indicating that it is appropriate to consider them as fully resolved DNS.
\begin{table}[t!]
    \centering
    \caption{Compressible fHIT operating conditions, grid size $N_p$, time step $\Delta t$, target forcing coefficient $\theta^\ast$, and resultant target mean fluctuation amplitude $\langle u^\prime u^\prime\rangle^{*}_{av}$, Taylor Reynolds number $\mathrm{Re}_\lambda$, and ratio of the grid size to the Kolmogorov microscale $\Delta x / \eta$.}\label{tab:fHIT_conditions}
    \begin{tabular}{lllll|llll}
        \toprule
         $\mathrm{Re}$ & $\mathrm{Ma}$  & $N_p$ & $\Delta t$ & $\theta^{*}$ & $\langle u^\prime u^\prime\rangle^{*}_{av}$ & $\mathrm{Re}_\lambda$ & $\Delta x / \eta$\\
         \midrule
         50 & 0.25 & $64^3$ & 0.01 & 1/32 & 0.3089 & 48 & 0.77\\
         120 & 0.25 & $256^3$ & 0.0025 & 1/32 & 0.3531 & 85 & 0.37\\
         \bottomrule
    \end{tabular}
\end{table}

Both Reynolds number cases use the same value of the solenoidal forcing coefficient, $\theta^{*} = 1/32$, to generate the target data. We simulate the Navier--Stokes equations using $\theta^{\ast}$ and time-volume-average the velocity fluctuations to generate the target data: 
\begin{equation}
    \langle u^\prime u^\prime\rangle^\ast_{av} = \frac{1}{V}\int_\Omega \frac{\langle u^\prime u^\prime\rangle + \langle v^\prime v^\prime\rangle + \langle w^\prime w^\prime\rangle}{3} dx.
\end{equation}
Here, $\lbrace u^\prime, v^\prime, w^\prime\rbrace^\top$ are the fluctuations from the Reynolds decomposition of $\lbrace u, v, w\rbrace^\top$~\eqref{eq:Reynolds}, and $V = (2\pi)^3$ is the volume of the computational domain.

The optimization goal is to use the target value $\langle u^\prime u^\prime\rangle^{*}_{av}$, generated using $\theta^*$, to recover the target forcing coefficient $\theta^*$.
The objective function is therefore
\begin{equation}\label{eq:fHIT_loss}
    J(\theta) = \frac{1}{V}\int_\Omega \Bigg(\Big\langle u^\prime u^\prime\Big\rangle_{av}(x) - \Big\langle u^\prime u^\prime\Big\rangle^{*}_{av}\Bigg)^2 dx,
\end{equation}
where
\begin{equation}
    \Big\langle u^\prime u^\prime\Big\rangle_{av}(x) =
    \frac{\langle u^\prime u^\prime\rangle + \langle v^\prime v^\prime\rangle + \langle w^\prime w^\prime\rangle}{3}.
\end{equation}
When implementing this, we make use of the fact that $\langle u\rangle_{av} = 0$ by definition of the forcing function and boundary conditions and replace $\langle u^\prime u^\prime\rangle_{av}$ by $\langle u u\rangle_{av}$.
All optimization runs start from an initial parameter value of $\theta = 1/160$ and $\varepsilon = 1/160$.

In the learning rate decay schedule \eqref{eq:LR}, $\alpha_1$  was set to decay the learning rate to 0.0025 over 1500 time units, and the product $\alpha_0\ t_{decay} = 1000$ was used to set the onset of decay for all cases.
As we only consider a single parameter, the online gradient flow without RMSprop is used as the optimizer for all cases. Groups of three simulation trajectories (for a single evaluation of the gradient estimator) are distributed over 1 GPU of the low Reynolds case and 8 GPUs for the higher Reynolds number case. 

\subsection{Optimization results}
A summary of the optimization case conditions and hyperparameters is presented in Tab.~\ref{tab:fHIT_stats}.
At the low Reynolds number condition, we consider minibatch sizes of $N_{mb}=1$, 8, and 64, EWMA length $M = 1000$, and three different initial learning rates $\alpha_0$.
We repeat all $\mathrm{Re} = 50$ and $N_{mb}=1$ cases with $M=5,000$ to investigate the effect of the EMWA length on the optimization at low $N_{mb}$ in a more complex system.
For the high Reynolds number cases, we consider $N_{mb}=1$, 8, and 64 for $M=4,000$ and  $\alpha_0 = 0.05$.

The optimization results in Tab.~\ref{tab:fHIT_stats} consist of the $\theta$-$\mathrm{RMSE^*}$ \eqref{eq:RMSE}, and the estimated value of the loss function $J(\theta)$ \eqref{eq:fHIT_loss}, both calculated over the last 500 time units of the optimization.

For the low Reynolds number case, we see that, with an appropriate combination of $M$ and $N_{mb}$, the optimization has been able to reduce the error in the parameter to less than 1\% for the least aggressive learning rate schedule.
Errors with a more aggressive schedule are higher, although we expect that these can be reduced with further decay of the learning rate from the $\alpha(t) = 0.0025$ at the nominal end of our optimization runs.

The high Reynolds case shows a similar trend with increasing minibatch size. (For this case, we consider $\alpha(t) = 0.005$ to be the nominal end of the optimization period.)
Both $N_{mb}=8$ and 64 achieve a parameter error in the region of 1\% and all cases achieve a multiple order of magnitude reduction in the loss function.
\begin{table}[t!]
    \centering
    \caption{Normalized  RMSE of the optimized parameter $\theta$ and loss function $J(\theta)$ for fHIT cases, estimated over the last 500 time units of the optimization, for simulations at different scaling Reynolds numbers and with different hyperparameters $N_{mb}$, $M$, and $\alpha_0$.}\label{tab:fHIT_stats}
    \vspace{11pt}
    \begin{tabular}{c|ccc|cc}
        \toprule
        $\mathrm{Re}$ & $N_{mb}$ & $M$ & $\alpha_0$ & $\mathrm{RMSE}^*(\theta)$ & $J(\theta)$\\
        \midrule
        \multirow{12}[5]{*}{50}&&&0.005 &  3.71\% & $2.99\times 10^{-4}$ \\
        &1 & 1000 &0.01 &  10.1\% & $1.33\times 10^{-4}$ \\
        &&&0.05 &  6.33\% & $1.14\times 10^{-4}$ \\
        \cmidrule(lr){2-6}
        &&&0.005 &  0.31\% & $1.55\times 10^{-4}$ \\
        &1&5000&0.01 &  1.57\% & $1.08\times 10^{-4}$ \\
        &&&0.05 &  5.18\% & $3.33\times 10^{-4}$ \\
        \cmidrule(lr){2-6}
        &&&0.005 &  0.93\% & $5.39\times 10^{-6}$ \\
        &8&1000&0.01 &  1.52\% & $4.74\times 10^{-6}$ \\
        &&&0.05 &  2.28\% & $6.07\times 10^{-6}$ \\
        \cmidrule(lr){2-6}
        &&&0.005 &  0.60\% & $7.05\times 10^{-8}$ \\
        &64&1000&0.01 &  0.676\% & $8.78\times 10^{-7}$ \\
        &&&0.05 &  1.14\% & $2.80\times 10^{-6}$ \\
        \midrule
        \multirow{3}[3]{*}{120} & 1 & 4000 & 0.05 & 3.06\% & $5.44\times10^{-4}$  \\
        \cmidrule(lr){2-6}
        & 8 & 4000 & 0.05 & 0.73\% & $3.37\times10^{-5}$\\
        \cmidrule(lr){2-6}
        & 64 & 4000 & 0.05 & 1.00\% & $5.88\times10^{-6}$\\
        \bottomrule
    \end{tabular}
\end{table}

\begin{figure}[t!]
    \centering
    \includegraphics[width=0.9\linewidth]{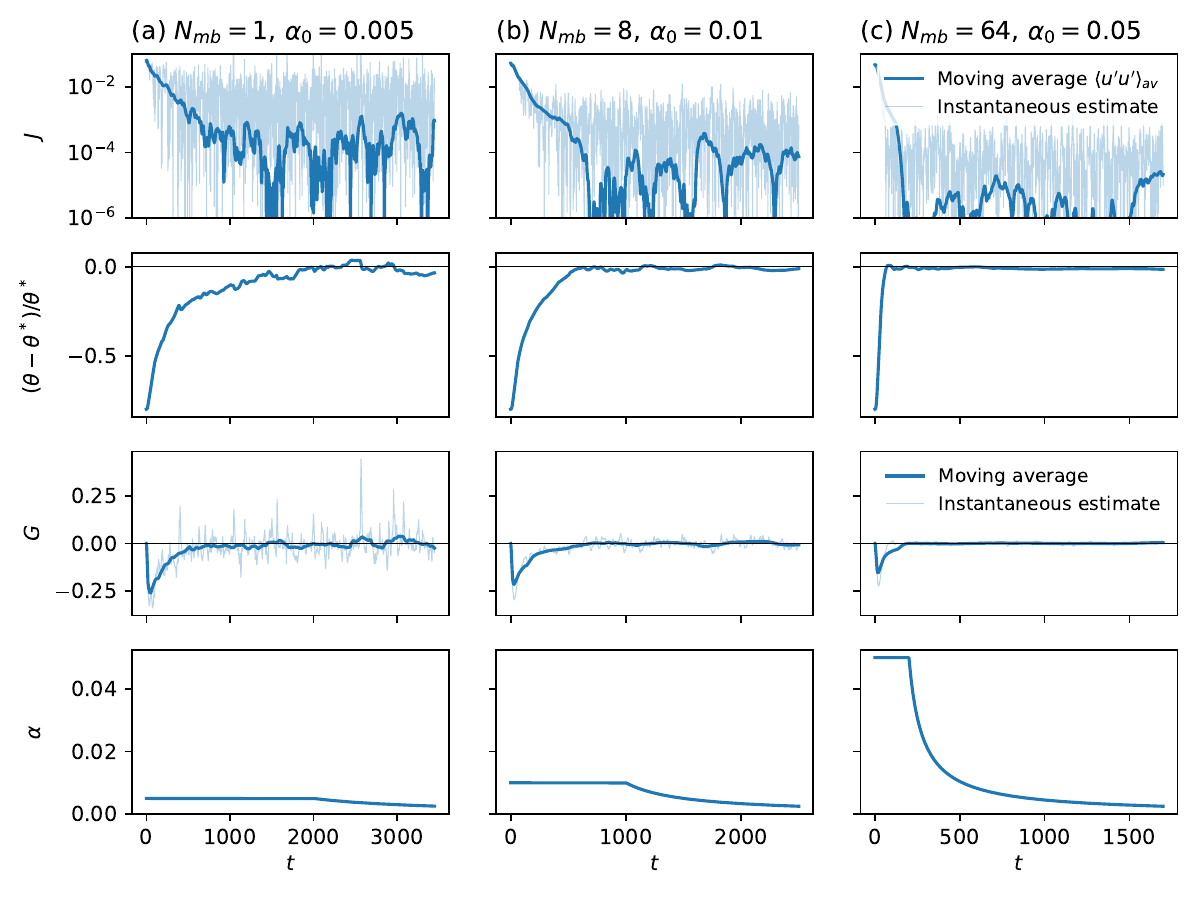}
    \caption{A subset of the $\mathrm{Re} = 50$ fHIT  optimizations (Tab.~\ref{tab:fHIT_stats}) with different minibatch sizes and learning rates. Estimated loss function $J(\theta)$, optimization parameter error $(\theta - \theta^{*})/\theta^{*}$, gradient estimate $G$, and learning rate $\alpha(t)$ with $M = 1,000$. Saturated lines for $J$ and $G$ correspond to $J$ and $\langle G \rangle$ estimated over 250 time units; light lines are their instantaneous values.
    }
    \label{fig:fHIT_general_results}
\end{figure}

\begin{figure}[t!]
    \centering
    \includegraphics[width=0.9\linewidth]{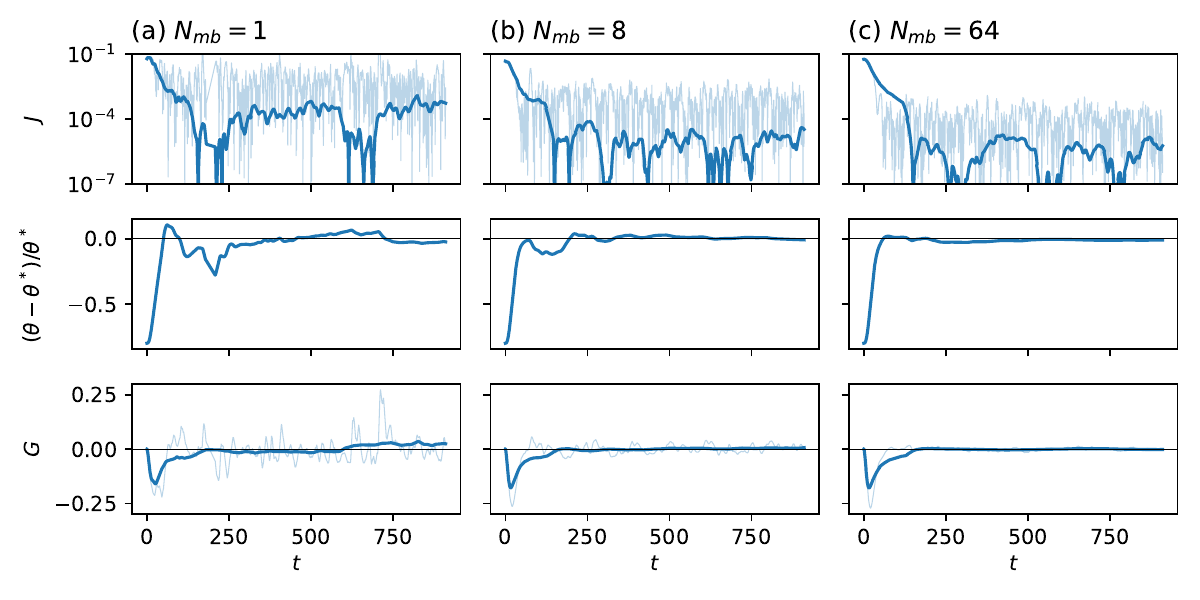}
    \caption{$\mathrm{Re} = 120$ fHIT optimization for different minibatch sizes. Estimated loss function $J(\theta)$, optimization parameter error $(\theta - \theta^{*})/\theta^{*}$, gradient estimate $G$, and learning rate $\alpha(t)$ with $M = 4000$. Saturated lines for $J$ and $G$ correspond to $J$ and $\langle G \rangle$ estimated over 250 time units; light lines are their instantaneous values.  Learning rate schedules are all the same as Fig.~\ref{fig:fHIT_general_results}c.}
    \label{fig:fHIT_highRe_results}
\end{figure}

Figures~\ref{fig:fHIT_general_results} and~\ref{fig:fHIT_highRe_results} show the convergence of the OGF-trained $\theta(t)$ to the correct value of $\theta^{\ast}$ for the low- and high-Reynolds number cases. These include a range of minibatch sizes (ranging from no minibatch, $N_{mb} = 1$, to $N_{mb} = 64$) and, for the low-Reynolds number case, initial learning rates.
In general, the larger the minibatch size, the less noisy the gradient estimator $G$  is, resulting in smoother convergence, and the faster the objective function $J(\theta(t))$ converges.
In particular, for $N_{mb} = 64$ the noise is barely visible, and the trained parameter $\theta(t)$ rapidly converges.
Note that in Fig.~\ref{fig:fHIT_general_results} we plot a subset of the cases in Tab.~\ref{tab:fHIT_stats}---this subset was chosen to provide an example of each different $\alpha_0$ and $N_{mb}$.

This section demonstrates that the OGF method can successfully optimize over the steady-state statistics of turbulent flows at scale.
We consider flows with up to $8.4\times10^7$ degrees of freedom per independent simulation trajectory. Even without minibatching, the method is able to converge to within 1\% of the optimal parameters.
Over the last three sections, we have presented results with a wide range of hyperparameters and systems. In the next section, we draw these together to extract general trends in the performance of the method across all three cases.

\section{Hyperparameter sensitivity}\label{sec:hyperparameter_sensitivity}

An important aspect of assessing any optimization methodology is its sensitivity to hyperparameters. If an optimization algorithm only converges for a small range of hyperparameter values, then it will be of limited practical use.
In general, we have found the OGF methodology to be robust to changes in the minibatch size, finite difference step size, optimizer, and learning rate schedule.

In this section, we synthesize results from the three test cases to explore the sensitivity of the method to its hyperparameters.
We first consider the minibatch size and EWMA length (both used to reduce the noise in the gradient estimator).
For this, we predominately focus on the trends we observe in the KSE and fHIT test cases---as PDEs, we expect their behavior to be most representative of potential future applications of the OGF method.
We then consider the finite-difference step size, after which we consider the influence of the choice of optimization algorithm.
Finally, we examine the effect of the initial learning rate on the convergence of the method for the fHIT case, which has the richest dynamics, and is consequently most likely (in our opinion) to display adverse effects from overly aggressive learning rate schedules.

\subsection{Minibatch size and EWMA length}

Some form of averaging is necessary in our algorithm, as any instantaneous gradient estimate contains noise. To reduce this, we use minibatching and EWMAs. EWMAs do not require additional computational resources but do introduce temporal correlations into the estimate. Minibatching with independent samples does not introduce any correlation (and thus is better for noise reduction), but scales linearly in the number of samples.

The ability of minibatching to reduce noise can be seen in Fig.~\ref{fig:fHIT_general_results} for fHIT at $Re=50$. Crucially, less noise improves the accuracy of the gradient estimator $G$ and thus the future parameters. The parameters in return determine the flow fields and future values of $G$. Without sufficient averaging, the algorithm requires much longer to converge (in conjunction with a smaller learning rate). In Fig. \ref{fig:KSE_comp_mb}a for the KSE, we see the convergence of the parameter $\nu$ stagnating due to insufficient averaging.
Figures~\ref{fig:fHIT_general_results} and~\ref{fig:fHIT_highRe_results} illustrate for the fHIT at $Re=50$ and $120$ that increasing the number of sample trajectories indeed results in smaller RMSEs.
Higher minibatch sizes further reduce the loss function estimate, as observed for all cases in Tabs.~\ref{tab:lorenz_J},~\ref{tab:KSE_J}, and~\ref{tab:fHIT_stats}.
While the superior parameter estimates might help with reducing $J$, we attribute this primarily to statistical noise reduction in $J$ itself.

The temporal averaging introduced by the EWMA can reduce noise in the gradient estimator and speed convergence (especially in cases with moderate minibatching), although there is a limit to how large the EWMA length $M$ can be before causing convergence difficulties.
For the low Reynolds number fHIT in the case without minibatching ($N_{mb}=1$ in Tab.~\ref{tab:fHIT_stats}) $M=5,000$ performs better than $M=1,000$ across all cases. For the KSE, we see the same trend of decreasing parameter error with increasing $M$ in Tabs.~\ref{tab:KSE_c_err} and~\ref{tab:KSE_nu_err} for $N_{mb} = 10$ up to a point $M\approx1,000$ where the error starts to slowly creep upwards again. 

\begin{figure}[t!]
    \centering
    \includegraphics[width=0.9\linewidth]{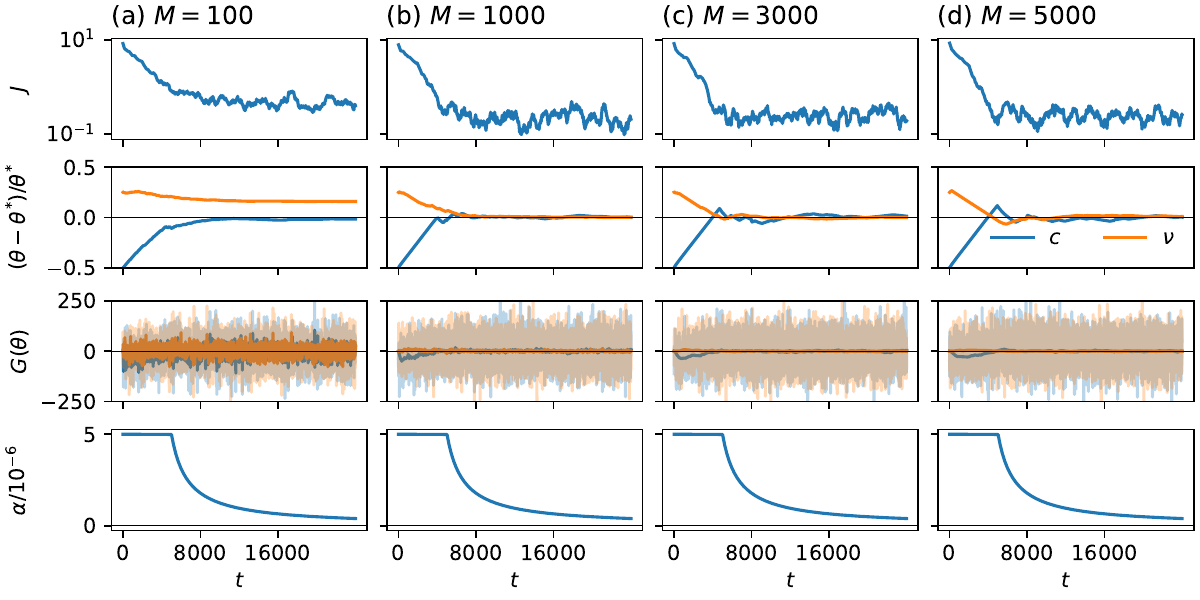}
    \caption{Comparison of different EWMA parameters $M$ at fixed minibatch size $N_{mb} = 10$. KSE loss function $J(\theta)$ estimated over $1,000$ time units, optimization parameter error $(\theta - \theta^{*})/\theta^{*}$, gradient estimate $G$, and learning rate $\alpha(t)$. Saturated lines correspond to the EWMA estimates, light lines are instantaneous values. } 
    \label{fig:KSE_comp_M}
\end{figure}

To explain this, we compare the optimization trajectories for different choices of $M$ in the KSE optimization (Fig.~\ref{fig:KSE_comp_M}). Let us focus separately on the initial transient and the (almost) steady "fine-tuning" phase. The benefit of using an EWMA is most visible during the transient phase. Here, the first column with $M=100$ shows that the parameters, in particular $\nu$, converge only very slowly to their target value. This illustrates that the transient phase is most prone to noise.
While the system exhibits a clear gradient signal, Fig.~\ref{fig:KSE_comp_M} indicates the EWMA gradient estimator is not able to capture this signal. The result is worse convergence compared to the cases with $M \geq 1,000$ (Fig.~\ref{fig:KSE_comp_M}b-d).

After the parameters reached a small environment around the target values, the instantaneous gradient estimates gain more importance. One potential issue of using the EWMA with a large $M$ manifests itself in over-/undershooting, as seen for the KSE with $M=5,000$ in Fig.~\ref{fig:KSE_comp_M}. If a parameter overshoots the target parameter, the instantaneous gradient (provided a sufficiently large minibatch) captures the overshoot and changes sign, but the EWMAs ability to adapt to a change in sign is limited for large $M$. This is analogous to the behavior of momentum in common machine learning optimizers. 
A possible stagnation in convergence is mitigated by decaying the learning rate over time, reducing the overshoot, and allowing the algorithm to converge more quickly.
Although we do not explore this, it is possible that decaying $M$ in conjunction with the learning rate could further help address the over-/undershooting.

A final point worth highlighting on minibatching and the EMWA is that for fHIT, we observe in Tab.~\ref{tab:fHIT_stats} that our algorithm achieves $O(1\%)$ RMSE for the forcing parameter. In particular, the results for $N_{mb} = 1$ and $M=5,000$ are encouraging, as they illustrate that a sufficiently large EWMA ensures convergence up to a small error while not requiring additional parallel resources.

\subsection{Finite difference step size}
\begin{figure}[t!]
    \centering
    \includegraphics[width=0.9\linewidth]{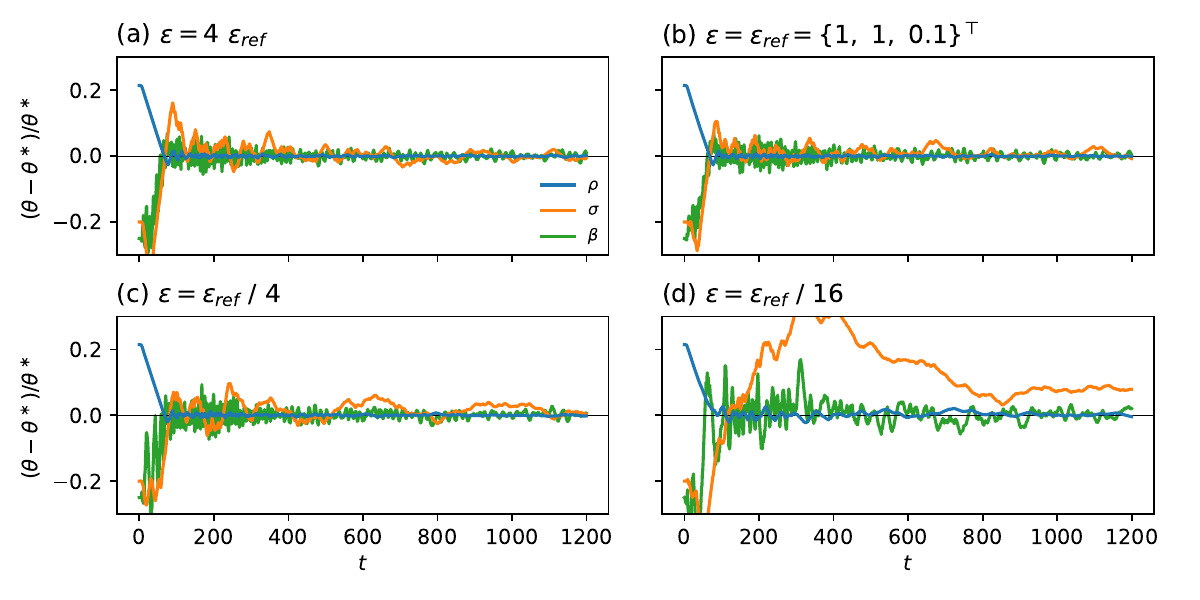}
    \vspace{-11pt}
    \caption{Convergence of the online gradient flow algorithm applied to the Lorenz-63 with different finite difference step sizes. All cases use $N_{mb} = 10$ and $M = 1,000$ and the same learning rate schedule as Fig.~\ref{fig:lorenz_general_plot}.}
    \label{fig:lorenz_epsilon}
\end{figure}

The finite difference step size $\bm\varepsilon$ is a hyperparameter that does not feature in adjoint optimization methods and most general machine learning algorithms.
When choosing $\bm\varepsilon$ in our proposed methodology, there is a tradeoff between a more accurate derivative with a small $\varepsilon$ and reduced noise in the gradient estimate with a larger $\varepsilon$.
Figure~\ref{fig:lorenz_epsilon} shows this for the Lorenz-63 system~\eqref{eq:lorenz63}.
The online gradient flow methodology converges for a wide range of finite difference step sizes (a factor of 4 either side of our nominal reference step size), all without changes to the learning rate schedule in Fig.~\ref{fig:lorenz_general_plot}.
Using a larger finite difference step size is expected to result in less noise in the gradient estimator and smoother convergence (although this is not clear from Fig.~\ref{fig:lorenz_epsilon}a).
With the current central difference approximation for $\pp{\langle u\rangle}{\btheta}$, the truncation error in the gradient estimate has a small prefactor of $\varepsilon^2$. In our experience, time-averaged loss functions tend to be fairly smooth, so we see that convergence of the methodology to the correct parameter values is unaffected by larger finite difference step sizes. 

Reducing the finite difference step size by a factor of four in Fig.~\ref{fig:lorenz_epsilon}c increases the noise in the optimal parameter estimates but the algorithm still achieves reasonable convergence.
Further reduction of the step size in Fig.~\ref{fig:lorenz_epsilon}d results in significant noise in the parameter estimates and issues converging the $\sigma$ parameter. We expect that the algorithm will still be capable of converging even with this small finite difference step size, but it will require smaller learning rates and longer run times.

\subsection{Choice of optimizer: SGD versus RMSprop}
\begin{figure}[t!]
    \centering
    \includegraphics[width=0.9\linewidth]{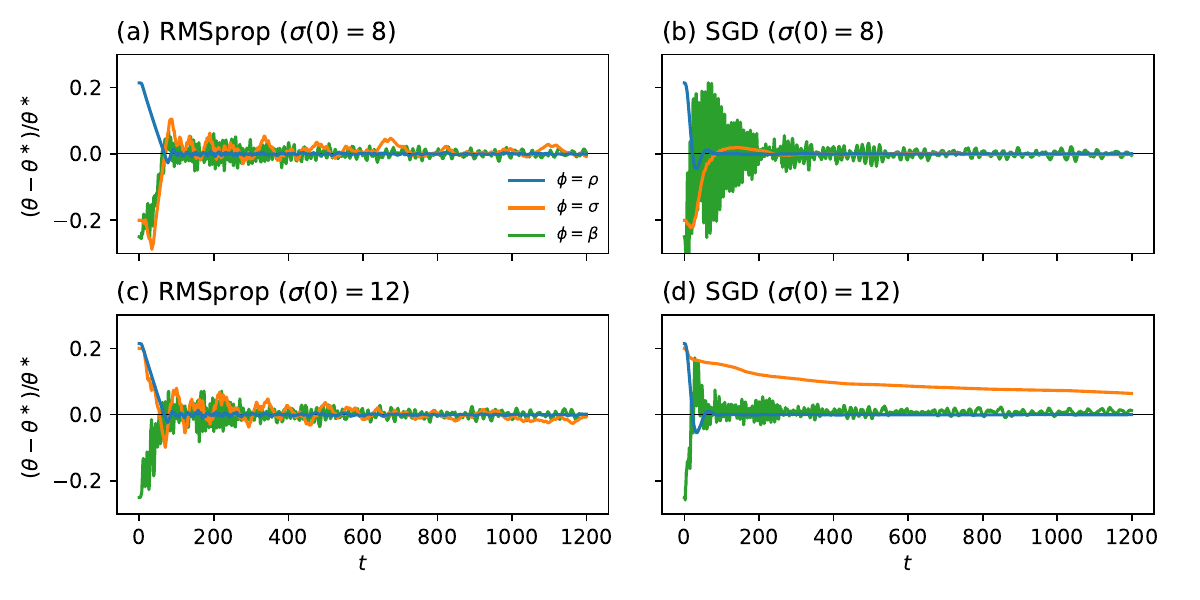}
    \vspace{-11pt}
    \caption{Convergence of the online gradient flow algorithm applied to the Lorenz-63 with different choices of optimizer (RMSprop or SGD) and initial values for $\sigma$. All cases use $N_{mb} = 10$ and $M = 1,000$.}
    \label{fig:lorenz_rms_prop}
\end{figure}

\begin{figure}[t!]
    \centering
    \includegraphics[width=0.6\linewidth]{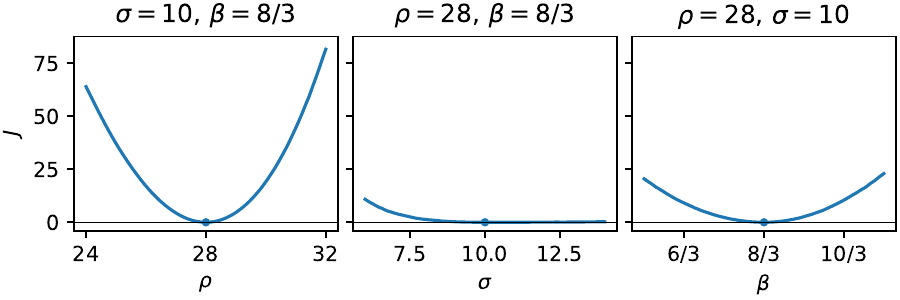}
    \vspace{-11pt}
    \caption{Projection of the objective function $J$ on to the parameter space in the neighborhood of its minima. Target values of the parameters are marked with a dot.}
    \label{fig:lorenz_params}
\end{figure}

It is interesting to compare the convergence of the algorithm with different optimizers.
We consider the Lorenz-63 system~\eqref{eq:lorenz63} with two different starting values of $\sigma(0)$, $\sigma(0) = 8$ (Figs.~\ref{fig:lorenz_rms_prop}a and~b) and $\sigma(0) = 12$ (Figs.~\ref{fig:lorenz_rms_prop}c and~d).
These are below and above the target value of $\sigma^{*} = 10$.
The learning rate---but not decay schedule---is adjusted for SGD to give qualitatively similar behavior to RMSprop.
Approaching the target value of $\sigma^{*}$ from below, we see that both RMSprop and SGD converge to the target parameter values, and SGD has slightly less noise in the parameter estimates due to the increase in the effective learning rate of RMSprop when $\widetilde{G^2} \approx 0$.
When approaching $\sigma^{*}$ from above, RMSprop has a similar convergence to before, but for SGD the convergence of $\sigma$ to $\sigma^{*}$ is much slower.
This can be explained by examining the loss function---we show projections of this in Fig.~\ref{fig:lorenz_params}.
$J$ is relatively well conditioned in the $\rho$ and $\beta$ directions.
In the $\sigma$ direction there is a large gradient pointing towards the target value of $\sigma^\ast = 10$ from below, but the loss function is very flat above $\sigma^\ast$.
RMSprop is very effective at accelerating convergence by modifying the learning rate appropriately the learning rate in low (or high) gradient regions.

\subsection{Learning rate}
The online methodology is able to converge very closely ($\mathcal{O}(1\%)$ RMSE error) to the optimal value of the parameters independent of the initial learning rate or learning rate schedule.
We make no claims that the learning rate schedules we have used are optimal for each case.
Our experience is that for a fixed final learning rate, larger initial learning rates and more aggressive learning rate decay schedules will result in more error due to history effects in the system.
However, we expect the error in all cases to continue to decrease with further optimization time and learning rate decay, so it may still be that for some cases high initial learning rates are more computationally efficient for achieving a given error.

As can be seen in Tab.~\ref{tab:fHIT_stats} for the $\mathrm{Re} = 50$ fHIT case,
the error in the parameters and estimated loss increase for increasing initial learning rate and increasingly aggressive learning rate decay schedule.
This is most clear for the larger minibatch sizes ($N_{mb} \geq 8$) with initial learning rates $\alpha_0 = 0.01$ to $0.05$.
A possible explanation for this is the history effects in the fHIT system: at the steady-state of the system, the energy being added to the smallest wavenumbers (largest scales) of the system is balanced by the dissipation at the largest wavenumbers (smallest scales).
However, when the forcing amplitude increases rapidly due to a large initial learning rate, we conjecture that energy builds up in the lowest wavenumbers and dilatational modes, and the system takes longer to equilibrate to its statistical steady-state.

\FloatBarrier
\section{Conclusion}
Optimizing over the statistical steady-state of chaotic dynamics is a challenging problem. Given the increasing relevance of turbulent flow simulations to engineering applications, the development of practical techniques for optimizing over the steady-state of chaotic dynamics  could have a broad impact.
However, there are currently no  methods that are capable of doing this in a scalable manner.
In particular, the chaotic nature of turbulence causes conventional adjoint techniques to fail, and existing remedies for this lead to biased gradient estimates and/or excessive computational cost for large degree of freedom systems.

We develop an online gradient flow (OGF) method for optimizing over the statistical steady-state of chaotic dynamics, including turbulent flows governed by the Navier-Stokes equations. The approach builds on a numerical method with proven convergence in stochastic differential equations~\cite{Wang2022,Wang2024} and leverages techniques from machine learning: minibatch averaging, momentum, and adaptive optimizers. The OGF method can be viewed as a form of stochastic gradient descent for chaotic dynamics. Although the its computational cost scales linearly with the numbers of parameters, the method can be perfectly parallelized (up to a small amount of communication) and---critically---does not have an adverse scaling with the degrees of freedom in the target simulation. The method is therefore computationally feasible for optimization over the statistical steady-state of large degree of freedom simulations of chaotic PDEs, of which turbulence resolving Navier--Stokes calculations are a prime example with real-world applications.

We apply the OGF method to a series of chaotic systems with increasing complexity and computational cost.
We evaluate the method's performance for the Lorenz-63 system of ODEs~\cite{Lorenz1963} and the Kuramoto--Sivashinsky PDE~\cite{Kuramoto-1976,Kuramoto1978,Sivashinsky1977}. Finally, we use the method to optimize over compressible forced homogeneous isotropic turbulence, with example simulations having up to $8.4\times10^7$ degrees of freedom, large enough to be representative of physically realistic simulations. For all cases, the method accurately and robustly recovers \emph{a priori} known optimal parameter values, usually achieving multiple orders of magnitude reductions in the loss function and parameter estimates, to within 1\% of their known optimal values. 

Due to its online nature, the OGF method continually updates the parameters in the estimated gradient descent direction. This is in contrast to offline methods, which only update the parameters once every optimization time interval. The lack of the optimization time interval as a hyperparameter in the OGF method is a second critical factor in the performance of the OGF method compared to offline gradient descent. In offline methods, this time interval must be specified \emph{a priori} and must be long enough to observe all physically relevant timescales, yet short enough to be computationally tractable for several iterations of the offline gradient estimate. This balance is not always possible to find, and where it is, requires a high degree of trial and error, which further increases the cost of offline optimization.

We have observed the OGF method to be relatively insensitive to variations in hyperparameters across the range of systems we test, with effects that are broadly intuitive to machine learning practitioners.
The method converges as long as sufficient minibatching and exponentially weighted moving averaging are applied, both of which reduce noise in the gradient estimate,
though excessive length of the moving average can lead to over/undershoots in the fine-tuning stage of the optimization process.
A particularly important aspect for the computational cost of the OGF method is its ability to converge to within 1\% of known optimal parameters using the exponentially weighted moving average without minibatching (on which the cost of the method depends linearly) two out of three of our examples (the Lorenz-63 system and forced homogeneous isotropic turbulence).
We also see that the method is insensitive to the finite-difference step size used in the gradient estimator and propose that its sensitivity to the initial learning rate is dominated by the history effects of the physical system being optimized rather than any fundamental limitations of the method.

The OGF method enables optimization over the steady-state statistics of large degree of freedom chaotic systems, with a particular focus on (but not limited to) unsteady turbulent flows.
To the authors' knowledge, there are currently no existing methods capable of doing this.
To rigorously assess the new methodology, we have restricted our attention to cases which have \emph{a priori} known optimal parameters, with target data extracted from an exact solution of the system. A natural next step is to examine situations in which the target statistics are not necessarily exactly recoverable by the system---this is of particular importance to the closure modeling of unresolved/unknown physics.

Given that the quantities of interest from these unsteady turbulent flow simulations (and experimental measurements from their physical counterparts) usually take the form of time-averaged statistics, the OGF method has the potential to significantly enhance the ability of engineers and scientists to directly optimize closure models, flow controllers, and flow geometries at a significantly reduced cost compared to existing methods.

\section*{Acknowledgments}
This work was supported by the U.S.\ Office of Naval Research under award N00014-22-1-2441. This material is based in part upon work supported by the U.S.\ National Science Foundation under Award CBET-22-15472. 
This work is supported by the U.K.\ Engineering and Physical Sciences Research Council grant EP/X031640/1.
This work used resources of the Oak Ridge Leadership Computing Facility, which is a DOE Office of Science User Facility supported under Contract DE-AC05-00OR22725.
The computations described in this research were performed using the Baskerville Tier 2 HPC service (https://www.baskerville.ac.uk/). Baskerville is funded by the EPSRC and UKRI through the World Class Labs scheme (EP/T022221/1) and the Digital Research Infrastructure programme (EP/W032244/1) and is operated by Advanced Research Computing at the University of Birmingham.
For the purpose of Open Access, the authors have applied a CC-BY public copyright license to any author accepted manuscript version arising from this submission.

\bibliographystyle{unsrtnat}
\bibliography{refs}

\end{document}